\documentclass[10pt]{article} 
\usepackage[table]{xcolor} 
\usepackage[accepted]{tmlr}

\usepackage{hyperref}
\usepackage{url}            
\usepackage{booktabs}       
\usepackage{amsfonts}       
\usepackage{nicefrac}       
\usepackage{microtype}      
\usepackage{graphicx}
\usepackage{booktabs}
\usepackage{algorithm}
\usepackage{algpseudocode}
\usepackage{enumitem}
\usepackage{wrapfig}

\usepackage{graphicx,amsfonts,amscd,amssymb,bm,url,color,latexsym,bbm,amsthm,amsmath}
\usepackage{physics}
\allowdisplaybreaks
\usepackage[capitalize,nameinlink]{cleveref}
\usepackage{hyperref}

\renewcommand{\mathbf}{\boldsymbol}

\newcommand{\mb}{\mathbf}

\newcommand{\bb}{\mathbb}

\newcommand{\paren}{\pqty}

\DeclareMathOperator{\st}{s.t.}

\usepackage{amsmath,amsfonts,bm}
\usepackage{ amssymb }

\def\eqref#1{equation~\ref{#1}}

\def\1{\bm{1}}

\DeclareMathAlphabet{\mathsfit}{\encodingdefault}{\sfdefault}{m}{sl}
\SetMathAlphabet{\mathsfit}{bold}{\encodingdefault}{\sfdefault}{bx}{n}

\def\gI{{\mathcal{I}}}

\usepackage{mathtools}

\definecolor{Gray}{gray}{0.92}
\usepackage{tabularx,arydshln}
\usepackage{tablefootnote}
\usepackage{sidecap}

\title{A Baseline Method for Removing Invisible Image Watermarks using Deep Image Prior}

\author{\name Hengyue Liang \email liang656@umn.edu \\
      \addr Electrical and Computer Engineering\\
      University of Minnesota, Twin Cities
      \AND
      \name Taihui Li \email lixx5027@umn.edu \\
      \addr Computer Science and Engineering\\
      University of Minnesota, Twin Cities
      \AND
      \name Ju Sun \email jusun@umn.edu\\
      \addr Computer Science and Engineering\\
      University of Minnesota, Twin Cities}

\def\month{07}  
\def\year{2025} 
\def\openreview{\url{https://openreview.net/forum?id=g85Vxlrq0O}} 

\begin{document}
\maketitle
\begin{abstract}
Image watermarks have been considered a promising technique to help detect AI-generated content, which can be used to protect copyright or prevent fake image abuse. In this work, we present a black-box method for removing \emph{invisible} image watermarks, without the need of any dataset of watermarked images or any knowledge about the watermark system. Our approach is simple to implement: given a \emph{single} watermarked image, we regress it by deep image prior (DIP). We show that from the intermediate steps of DIP one can reliably find an evasion image that can remove invisible watermarks while preserving high image quality. Due to its unique working mechanism and practical effectiveness, we advocate including DIP as a baseline invasion method for benchmarking the robustness of watermarking systems. Finally, by showing the limited ability of DIP and other existing black-box methods in evading training-based \emph{visible} watermarks, we discuss the positive implications on the practical use of training-based \emph{visible} watermarks to prevent misinformation abuse. 
Our code is publicly available at: \url{https://github.com/sun-umn/DIP_Watermark_Evasion_TMLR}. 
\end{abstract}

\section{Introduction}
\label{Sec: Intro}
In this prosperous era of generative AI, the traceability of AI-generated content (e.g., language, images, and videos) to its source has been frequently mentioned as a promising solution to promote the responsible use of generative AI \citep{fan2023trustworthiness}, e.g., to protect copyright or to curb misinformation. In particular, the traceability of AI-generated images has become increasingly urgent, as many AI products, such as DALL-E \citep{ramesh2022hierarchical} and Stable Diffusion \citep{rombach2022high}, can create highly photorealistic and artistic images that are hard to distinguish from natural photos or human drawings. Unsurprisingly, some AI-generated images have caused false beliefs on social media \citep{bbc2024ai}. Thus, major tech companies such as Google, OpenAI \citep{rueters2023openai} have recently opted to incorporate watermarks into their image generation products to improve traceability and promote responsible use.

\begin{figure*}[!htbp]
\centering
\begin{tabular}{ccc}
\includegraphics[width=0.23\textwidth]{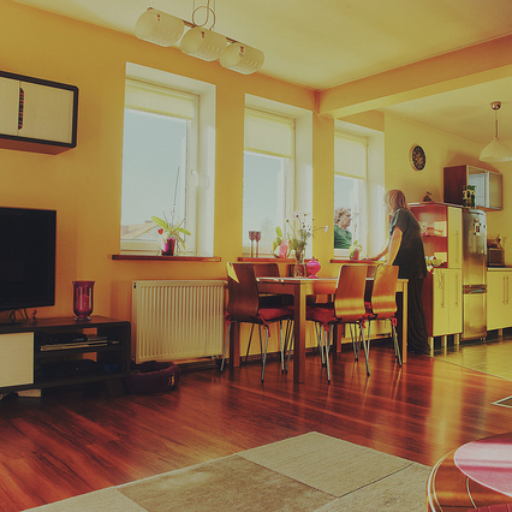}
&\includegraphics[width=0.23\textwidth]{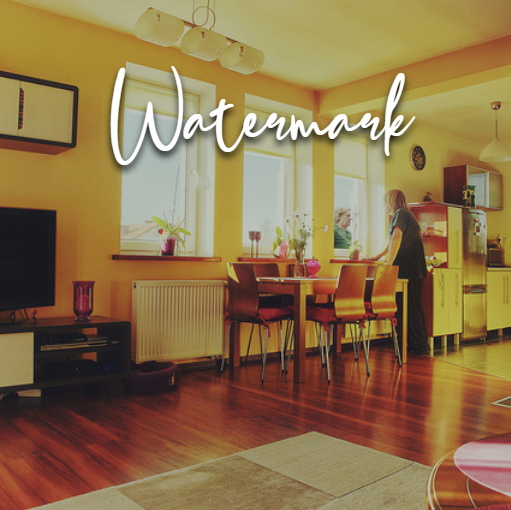}
&\includegraphics[width=0.23\textwidth]{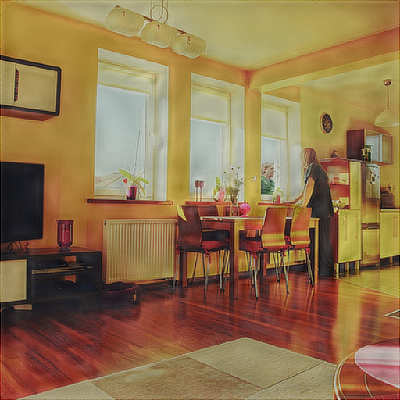}\\
\small{\textbf{(a)}}%
&\small{\textbf{(b)}}%
&\small{\textbf{(c)}}%
\end{tabular}
\caption{Example of (a) a clean image, (b) an image with an overlaid logo watermark and (c) an image with steganography watermark.}%
\vspace{-0.5em}
\label{Fig: watermark vis intro}
\end{figure*}
\begin{figure*}[!htbp]
\centering
\includegraphics[width=0.75\textwidth]{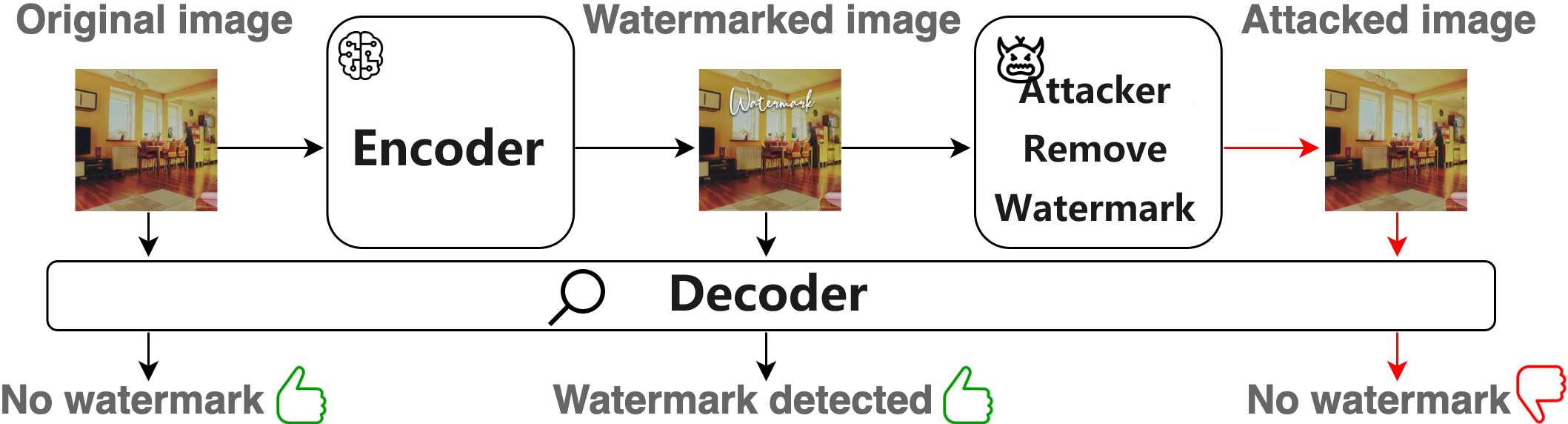}
\caption{Illustration of the watermarking system and the robustness concern---watermarks may be removed with little loss of image quality.}%
\vspace{-1em}
\label{Fig: watermark system vis}
\end{figure*}

\paragraph{Manually designed vs. training-based watermarks}
Watermarking methods can be generally divided into two categories: \textbf{(i)} \emph{non-blind} methods \citep{cox1997secure,hsieh2001hiding,pereira2000fast} and \textbf{(ii)} \emph{blind} methods \citep{bi2007robust}, which are divided by whether access to original clean images is required to correctly decode the watermarked images~\citep{Zhao2024SoKWF}. In what follows, we focus only on blind methods (and refer to them as \emph{watermarks}), as they do not require access to clean images and fit better in large-scale application scenarios, such as tracing AI-generated content. Watermarks are typically embedded in images in terms of an overlaid logo or steganography\footnote{Steganography: detectable messages embedded in an image but are invisible to human eyes.} \citep{morkel2005overview}; see \cref{Fig: watermark vis intro} for an example. The embedded watermarks then are typically detected by human eyes or by algorithmic decoders \citep{voyatzis1999protecting,zhu2018hidden}. In early research, various manually designed watermarks were proposed and applied to copyright protection, e.g., visible watermarks (detectable by human eyes) such as an overlaid layer \citep{kankanhalli1999adaptive} or a color code signature (such as the example in \cref{App: DALLE-2 Example}); invisible watermarks (detectable by algorithmic decoders) such as \cite{tirkel1993electronic,pereira2000robust,navas2008dwt}. However, these manually designed watermarks are challenged by robustness concerns, i.e., imperceptible corruption, digital editing, or deliberate attacks to the watermarked image can make them undetectable~\citep{remove2022remove,zhao2024invisible}; see also \cref{Fig: watermark system vis}. To address these challenges, recent work has shifted to training-based watermarking systems using deep neural networks (DNNs). By incorporating ideas from data augmentation \citep{mumuni2022data} and adversarial training \citep{goodfellow2014explaining}
, the robustness of training-based watermarks against common digital corruptions consistently beats that of manually designed ones~\citep{zhu2018hidden,zhang2019robust,tancik2020stegastamp,jia2021mbrs}. 

\paragraph{Evaluating the robustness of watermark systems} 
Robustness evaluation of watermark systems often operates under either the white-box or the black-box evasion (i.e., attack) setting~\citep{Zhao2024SoKWF,an2024benchmarking}. In 
white-box evasion, the decoder and the groundtruth watermark are accessible to a dedicated evader~\citep{jiang2023evading}, which easily leads to a high threat. In contrast, the threat can be greatly reduced by maintaining secrecy of the system \citep{an2024benchmarking}, i.e., under the \emph{black-box} model, where the decoder and the groundtruth watermark remain unknown to the evader. Nowadays, it is common for companies not to open-source their generative products, e.g., DALLE-3 from OpenAI, Midjourney, Imagen from Google. Thus, being robust against potential black-box evasions is of higher priority in practical scenarios. 

The main focus of this paper is to explore \emph{black-box} evasion techniques that can generate evasion images with the best possible quality. We stress the image quality alongside the evasion success, because for a dedicated evader, the higher the quality of the evasion images they can produce, the more benefit they can potentially gain, e.g., breaking the copyright protection without compromising the image quality or misleading people to use highly naturally-looking fake images. Recently, researchers are putting effort into standardizing the robustness benchmark of image watermarks \citep{an2024benchmarking}. Although the selected black-box evasion methods are effective in bypassing watermark detection, their evasion images can still introduce visible defects; see \cref{Sec: exp and results} and \cref{Fig: watermark evasion vis stegaStamp}. Thus, the search for new methods to provide stronger stress tests of watermarking systems is still a pressing problem.

\paragraph{Our contributions} 
In this paper, we propose a new black-box watermark evasion method using the Deep Image Prior (DIP) \citep{ulyanov2018deep}---an untrained DNN-based prior that proves powerful in solving single-image blind denoising and numerous single-instance inverse problems~\citep{DBLP_journals_pami_QayyumISBBQ23,Tirer2023DeepIL,zhuang2023advancing}. Our main contributions include: \textbf{(i)} We show that DIP-based blind denoising can be used to generate high-quality evasion images effective against many existing invisible watermarks, both training-based and manually designed; \textbf{(ii)} We elucidate the principle behind DIP's evasion performance---its faster rate in picking up low frequencies than high ones empowers its image-agnostic watermark purification ability. Due to \textbf{(i)} and \textbf{(ii)}, we advocate including DIP evasion as an integral component in the robustness evaluation of watermarking systems~\citep{saberi2023robustness,an2024benchmarking}; \textbf{(iii)} Based on our analysis, we further recommend that to counteract black-box watermark evasions, a reliable watermark scheme should focus on modifying low-frequency components and have a reasonable magnitude. 
\section{Background and related work}
\label{Sec: background}

\begin{figure}[!tb]
\centering
\resizebox{0.95\linewidth}{!}{%
\includegraphics[width=\textwidth]{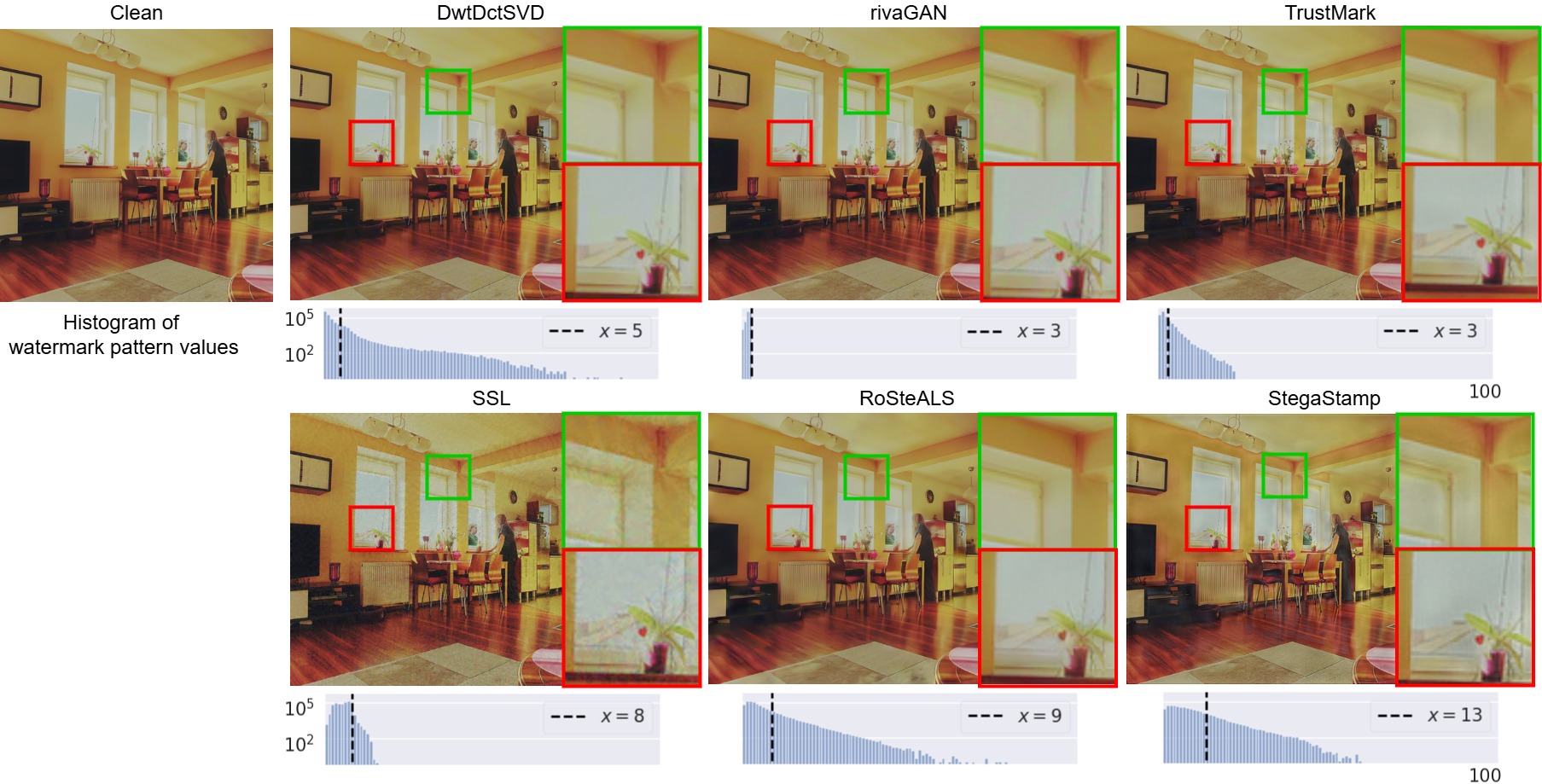}%
}
\vspace{-1em}
\caption{Visualization of a clean image (from COCO) and its steganography versions by different methods. Below each steganography image is the histogram of the pixel magnitude in the corresponding distortion (i.e., difference between the steganography image and the clean image): $y$-axis is in $\log$ scale and the vertical dashed line marks the $90\%$ quantile. While DwtDctSVD, rivaGAN and TrustMark induce distortions that are almost invisible, SSL, RoSteALS and StegaStamp produce relatively more visible distortions---local color jitters for SSL, global color jitters for StegaStamp, and global smoothing for RoSteALS. All images presented above are in \texttt{uint8} format with value in $[0, 255]$ with resolution $512 \times 512$.}%
\vspace{-0.5em}
\label{Fig: watermark vis}
\end{figure}

\paragraph{(Blind) image steganography} refers to the technique of hiding secret but retrievable messages in an image with minimal change to the image~\citep{zhu2018hidden}. Given an arbitrary natural image $I \in \gI$, where $\gI$ denotes the set of natural images, and an arbitrary $n$-bit message $\mb w \in \{0, 1\}^n$, an image steganography system typically consists of an encoder $E$---which takes any image $I$ and any message $\mb w$ and produces an encoded image, a decoder $D$---which takes any image and produces an informative message, and its system goal: ($\circ$ means function composition)  
\begin{subequations}
\label{eq:goal-stega}
\begin{align}
\label{eq: steganography def 1}
& (D \circ E) (I, \mb w) = \mb w, \; & \forall I \in \gI, & \; \forall \mb w \in \{0, 1\}^n, & \quad \text{(correctly encode and decode $\mb w$)} \\
\label{eq: steganography def 2}
& D(I) = \emptyset, \; & \forall I \in \gI, & & \quad \text{(no useful message decoded from a clean image)}\\
\label{eq: steganography def 3}
& E(I, \mb w) \approx I, \; & \forall I \in \gI,
 & \; \forall \mb w \in \{0, 1\}^n. & \quad \text{(minimal encoding distortion to the image)} 
\end{align}
\end{subequations}
Existing steganography methods differ by whether the encoder and decoder are manually designed or learned from data. Manually designed encoder-decoder pairs rely on ideas such as manipulating the least significant bit (LSB) \citep{tirkel1993electronic}, template matching in the Fourier domain \citep{pereira2000robust}, discrete wavelet transform (DWT), discrete cosine transform (DCT), and singular value decomposition (SVD) \citep{bi2007robust, pereira2000robust, navas2008dwt}. In contrast, training-based methods often learn DNN-based encoder-decoder pairs from data, based on variants of a model formulation derived from the goal stated in \cref{eq:goal-stega}: 
\begin{gather}
\begin{aligned}
\min_{\mb \phi, \mb \theta} ~ & \bb E_{\mb w, I} \ell_{m}[\mb w, (D_{\mb \theta} \circ E_{\mb \phi}) (I, \mb w)] & \text{(to ensure \cref{eq: steganography def 1})} \\
\st \; & \ell_{q} (I, E_{\mb \phi} (I, \mb w)) \leq \delta,  \quad \forall I \in \gI, ~ \forall \mb w \in \{0, 1\}^n, ~~ & \text{(to ensure \cref{eq: steganography def 3})}
\label{eq: learning-based steganography obj}
\end{aligned}   
\end{gather}
where $\mb \phi$ and $\mb \theta$ are learnable weights of the DNNs in $E$ and $D$, respectively; $\ell_m$ and $\ell_q$ are two losses measuring the error of \textbf{m}essage recovery and the \textbf{q}uality distortion to the image, respectively; and $\delta$ is the maximally allowed perturbation to the image caused by watermarking embedding. Representative training-based methods include HIDDEN and its variants \citep{zhu2018hidden,wen2019romark,luo2020distortion}, SteganoGAN \citep{zhang2019steganogan}, Stable Signature \citep{fernandez2023stable}, rivaGAN \citep{zhang2019robust}, StegaStamp \citep{tancik2020stegastamp}, Mbrs \citep{jia2021mbrs} and TrustMark \citep{bui2023trustmark}. These methods typically also incorporate regularization terms to encourage the distribution of the encoded images to be close to that of the original images based on generative adversarial networks (GAN) \citep{goodfellow2014generative}. SSL \citep{fernandez2022watermarking} and RoSteALS \citep{bui2023rosteals} are similar in spirit but perform learning in different spaces. In theory, solving \cref{eq: learning-based steganography obj} with a reasonably small $\delta$ can always produce distortion patterns that are \emph{invisible} to human eyes. However, existing methods typically work with heuristic penalty or regularization forms of \cref{eq: learning-based steganography obj} and therefore do not necessarily find feasible solutions to \cref{eq: learning-based steganography obj}, e.g., \citep{zhu2018hidden}. In addition, the choice of $\ell_q$ can differ, e.g., mean squared error (MSE)~\citep{zhu2018hidden,zhang2019steganogan}, LPIPS distance~\citep{zhang2018unreasonable}. Therefore, different methods lead to different levels of visible distortions. \cref{Fig: watermark vis} visualizes several popular training-based methods (together with the manually designed DwtDctSVD) and highlights the different distortion levels that they lead to.



\paragraph{(Blind) image watermarking}
Based on steganography, a natural way to trace AI-generated images is to assign \emph{a fixed message} $\mb w$ as the signature of the content owner (e.g., a company) and apply steganography to generate images containing the signature, i.e., watermarked images, to achieve~\citep{jiang2023evading}:
\begin{subequations}
\label{eq: watermark def}
\begin{align}
& D(E(I, \mb w)) = \mb w, \; & \forall I \in \gI, & \quad \text{($\mb w$ is a fixed message representing the signature)} \\
& D(I) \ne \mb w, \; & \forall I \in \gI, & \quad \text{(messages decoded from unwatermarked images should not be $\mb w$)}  \\
& E(I, \mb w) \approx I, \; & \forall I \in \gI.
& \quad \text{(minimal encoding distortion to the image)} 
\end{align}   
\end{subequations}
In practice, whether an image $I$ is watermarked by $\mb w$ can be detected by comparing the decoded message with $\mb w$:
\begin{equation}
\label{Eq: detection bar}
    \mathbbm{1} [ BA(D(I), \mb w) > \gamma],
\end{equation} where $BA$ denotes the bitwise accuracy and $\gamma$ is a preset task-dependent threshold~\citep{jiang2023evading,fernandez2023stable,yu2021artificial}.  Due to the similarity between \cref{eq: watermark def} and \cref{eq:goal-stega}, most existing work considers image watermarking as a special application of steganography, e.g., \cite{zhu2018hidden,tancik2020stegastamp,an2024benchmarking,Zhao2024SoKWF}. As a result, the learning formulation in \cref{eq: learning-based steganography obj} is also widely adopted in works that only focus on watermarking systems, e.g., \cite{zhang2019robust,fernandez2023stable}. 

\begin{wrapfigure}{r}{0.55\textwidth}
    \includegraphics[width=0.54\textwidth]{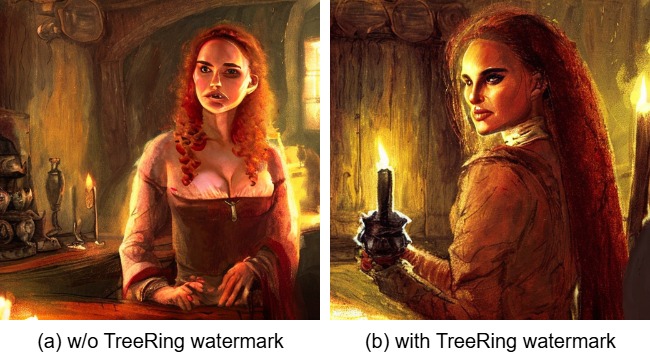}%
    \vspace{-1em}
    \caption{Visualization of a TreeRing watermark example, where \textbf{(a)} and \textbf{(b)} are images generated from the same text prompt input using the same image diffusion model without and with the in-process watermarking, respectively.}%
    \label{Fig: watermark vis treering}
    \vspace{-0.5em}
\end{wrapfigure}

The above watermark methods are \emph{post-processing} in nature, as the watermark is embedded on any given image $I$ that is already generated. There is an emerging line of \emph{in-processing} watermark methods that directly modify the image generation process~\citep{Zhao2024SoKWF,an2024benchmarking}, including TreeRing watermark~\citep{wen2023tree}, stable signature~\citep{fernandez2023stable}, Gaussian shading watermark~\citep{Yang2024GaussianSP}, and pseudorandom error-correcting code watermark~\cite{Gunn2024AnUW}. In these in-processing methods, 
there is no notion of ``clean images'' $I$ and every generated watermark image exhibits a semantic shift compared to the image generated without the in-process watermarking; see \cref{Fig: watermark vis treering} for an example generated by the TreeRing watermark. \textbf{In this paper, we focus on post-processing watermark methods due to their flexibility, as they are agnostic to the image generation process}.

\paragraph{Robustness of watermarking systems}
Robustness of an image watermarking system refers to the extent of the watermark to remain detectable by the decoder $D$ when the watermarked image is manipulated (also called ``evaded'' if such manipulation is a deliberate attack). Thus, robustness is typically associated with the potential of a watermarking system to be applied to copyright protection and misinformation detection. To stress test the robustness of watermark systems, various watermark evasion techniques have been proposed. These techniques are broadly classified into white-box and black-box evasions, depending on whether any component of the watermark system is known to the evader~\citep{an2024benchmarking}. \textbf{In this paper, we focus on black-box evasions where nothing about the watermark system is known}, as in practice, companies tend to keep their watermark system private. Existing black-box evasions can be classified into two groups: \textbf{Corruption methods} try to distort watermarked images so that watermarks become corrupted and undetectable. Classical digital editing (e.g., applying Gaussian noise, Gaussian blur, JPEG compression, etc. \citep{voyatzis1999protecting}), the query-based adversarial attack (WevadeBQ) in \cite{jiang2023evading} and the surrogate attack in \cite{saberi2023robustness} belong to this group; \textbf{Purification methods} treat embedded watermark patterns as noise signals and attempt to remove them using denoising and regeneration techniques, such as BM3D \citep{dabov2007image}, diffusion models \citep{saberi2023robustness,zhao2024invisible} and VAE \citep{zhao2024invisible}. The rationale behind the purification methods is rooted in \cref{eq: watermark def}, where the original image, if recovered successfully, is always an evasion. In addition, the original image will be the ultimate threat to any watermarking system---the watermark is evaded by an image without any loss of quality.

\section{Our method: DIP for black-box watermark evasion}
\label{Sec: methods}

\subsection{Watermark evasion via DIP-based blind denoising}
\paragraph{Deep Image Prior (DIP)}
\label{subsec: DIP background} refers to the technique of using \textbf{untrained} DNN as an implicit prior for natural images in solving image recovery problems, \textbf{without training on massive data}: for any natural image $I$, DIP parametrizes it as $I = G_{\mb \theta}(\mb z)$, where $G_{\mb \theta}$ is typically a trainable convolutional neural network (CNN) and $\mb z$ is a fixed input (typically randomly drawn). Now consider the canonical optimization-based formulation for image recovery problems: 
\begin{align}
    \min\nolimits_{I} \; \ell\paren{\mb y, f(I)} + \lambda R(I),  
\end{align}
where $\mb y \approx f(I)$ is the observation model, $\ell(\mb y, f(I))$ measures the recovery error, and $R(\cdot)$ denotes regularization on $I$. DIP transforms the formulation into 
\begin{align}
      \min\nolimits_{\mb \theta} \; \ell\paren{\mb y, f(G_{\mb \theta}(\mb z))} + \lambda R(G_{\mb \theta}(\mb z)). 
\end{align}
Note that the optimization is \emph{only} with respect to the CNN weights $\mb \theta$. The resulting formulation is then solved by first-order optimizers, such as ADAM~\citep{kingma2014adam}. Such a simple strategy, in combination with appropriate early stopping methods~\citep{li2021self,wang2021early,shi2022measuring} that pick the best intermediate recovered images, has proved highly successful in solving a wide range of image recovery problems, from simple denoising~\citep{ulyanov2018deep} to advanced scientific and medical reconstruction problems~\citep{Tirer2023DeepIL,DBLP_journals_pami_QayyumISBBQ23,zhuang2023advancing,zhuang2024blind,zhuang_practical_2023,jdd_doubledip,li_random_2023,li2023deep}.

\begin{wrapfigure}{r}{0.4\textwidth}
    \vspace{-1em}
    \includegraphics[width=0.4\textwidth]{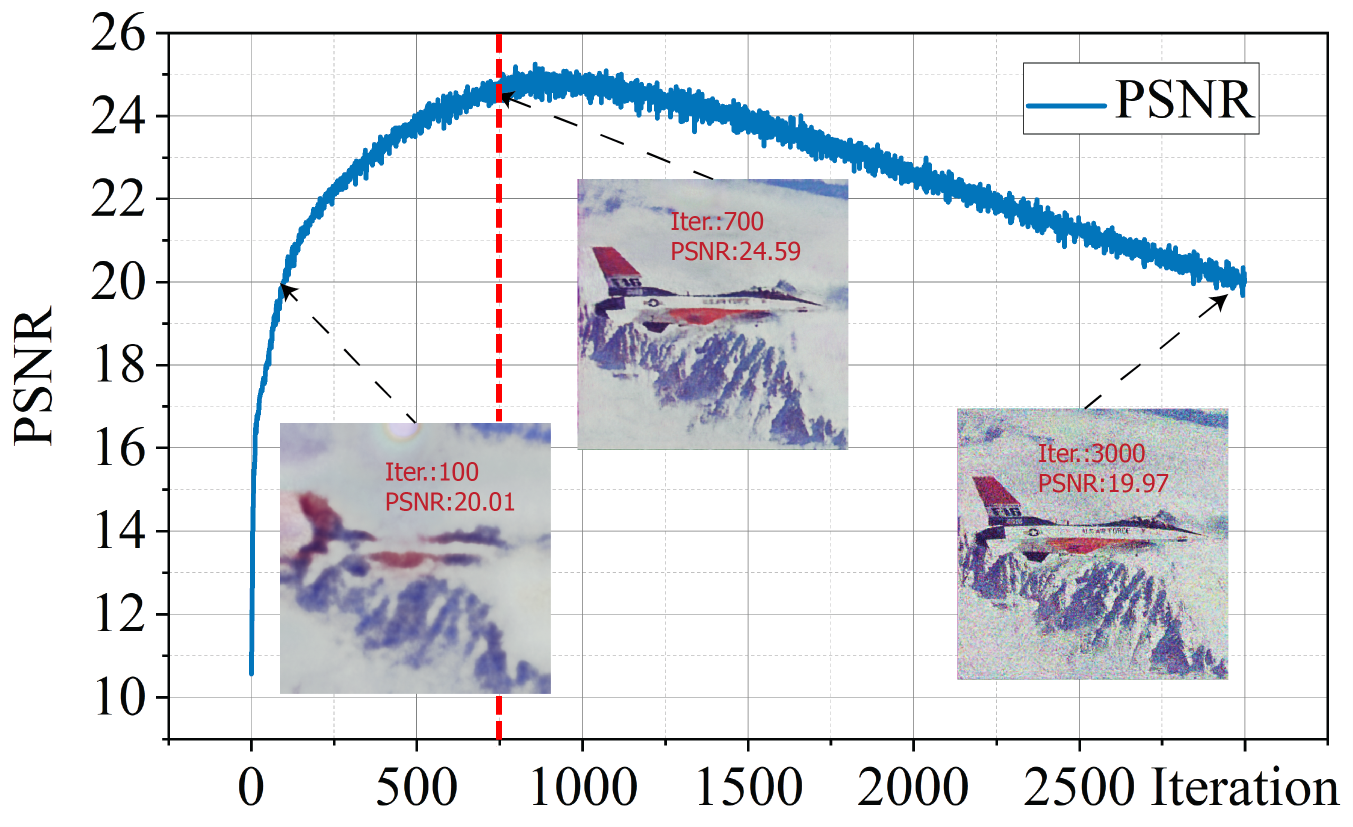}
    \vspace{-2em}
    \caption{A typical image quality (measured by peak signal-to-noise ratio, PSNR) vs. iteration curve when DIP is used in blind denoising tasks. An early stopping method is used to detect the iteration achieving the peak performance (red dashed line). (Figure adapted from \cite{wang2021early} under the Creative Commons 4.0 license)}
    \label{Fig: DIP demo}
    \vspace{-1em}
\end{wrapfigure}
\paragraph{Watermark evasion via DIP-based blind denoising}
\begin{algorithm}[tb]
\caption{DIP-based watermark evasion}
\label{alg: DIP evasion}
\begin{algorithmic}[1]
\Require A single watermarked image $I_{\mb w}$; a CNN $G_{\mb \theta} (\mb z)$ where $\mb \theta$ collects the trainable parameters and $\mb z$ is a fixed vector (randomly drawn as iid Gaussian); a watermark detection API whose input can be an arbitrary image and returns ``Yes/No'' as the detection output. 
\State Randomly initialize $\mb \theta^{(0)}$
\State Solve $\min_{\mb \theta} \ell_2 (I_{\mb w}, G_{\mb \theta}(\mb z))$ using the ADAM optimizer for a sufficient number of iterations $N$; record all intermediate results $I_i = G_{\mb \theta^{(i)}}(\mb z), ~ \forall i \leq N$, where $i$ denotes the iteration number.
\State Query the detection API using DIP intermediate steps ($I_i$'s); return the $I_i$'s that have no watermark detected.
\end{algorithmic}%
\end{algorithm}
Now, consider an arbitrary watermarked image $I_{\mb w} \doteq E(I, \mb w)$, where we do not know $E$ or $\mb w$. Since $I_{\mb w} \approx I$ as required in \cref{eq: watermark def} and $I$ is clearly a successful invasion, it is sensible to try to ``purify'' or ``denoise'' $I_{\mb w}$ toward $I$---the intuition behind all purification methods for black-box evasion~\citep{dabov2007image,saberi2023robustness,zhao2024invisible}. 

DIP has proven effective in single-image denoising, e.g., with Gaussian, impulse, shot noise, etc., when combined with appropriate early stopping strategies~\citep{mataev2019deepred,DBLP_conf_iccv_JoCC21,li2021self,wang2021early,li2023deep,li_random_2023,jdd_doubledip}. In particular, when the noise level is low---which is true for typical watermarking systems as $I_{\mb w}$ is supposed to be very close to $I$, a simple formulation with the standard mean-squared-error (MSE) loss and the additive noise model can perform ``blind'' simultaneous denoising for multiple types of noise~\cite{li2021self,wang2021early}. Inspired by this, we propose a simple DIP-based blind watermark-evasion formulation 
\begin{align} \label{eq:main-form}
    \min\nolimits_{\mb \theta} \; \norm{I_{\mb w} - G_{\mb \theta}(\mb z)}_2^2,  \quad (\textbf{DIP-based watermark evasion})
\end{align}
for any given watermarked image $I_{\mb w}$. Unlike DIP-based blind denoising which requires appropriate early stopping strategies to find optimal denoising, we only need to check the evasion success of all iterates when iteratively solving \cref{eq:main-form} by querying the watermark decoder. Our entire algorithm pipeline is summarized in \cref{alg: DIP evasion}. Although we do not invent DIP-based blind denoising, we are the first to explore it for blind evasion of invisible watermarks. While \citet{braindotai2021watermark} also performs DIP-based watermark removal, it aims at \emph{a different setup}: \citet{braindotai2021watermark} tries to remove visible watermarks by DIP-based image inpainting, which requires the \emph{exact} watermark location as its inpainting mask; in contrast, here we try to remove invisible watermarks using DIP-based denoising, which requires only the watermarked image and nothing else. 

\subsection{Why including DIP-based evasion as a baseline method? }
\label{sec:why-include-dip}
\begin{SCfigure}[][!htbp]
    \includegraphics[width=0.65\textwidth]{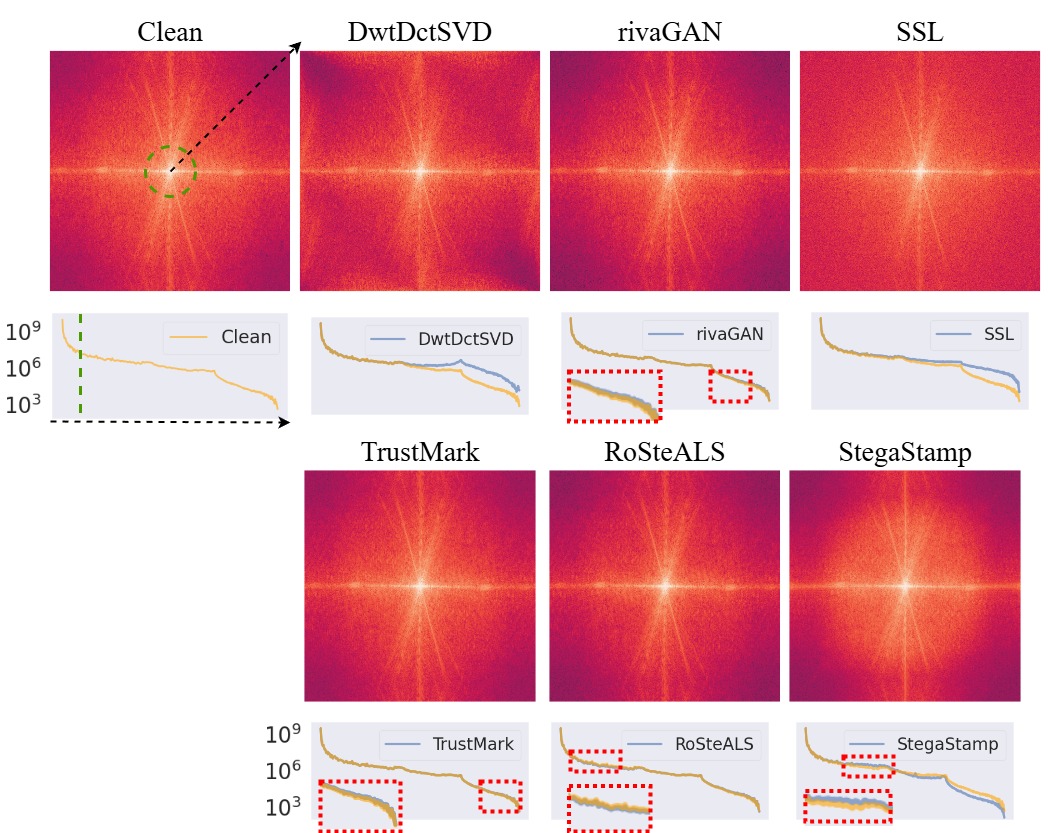}
    \caption{Visualization of 2D Fourier spectra of the clean and watermarked images from \cref{Fig: watermark vis} (magnitudes visualized in $\log$ scale). The histogram below each spectrum plot shows the band-wise energy distribution in the radial direction (i.e., the dashed arrow direction), where the $y$-axis is in $\log$ scale.}
    \label{Fig: watermark fourier vis}
    \vspace{-1em}
\end{SCfigure}
The primary obstacle for purification methods is that the ``noise'' patterns induced by watermark methods, especially those training-based ones, may not follow any simple noise model. Consequently, classical denoising methods that target specific noise types, such as BM3D, may struggle to remove the watermark. The recent regeneration techniques via diffusion models or VAE effectively perform learned denoising. However, a priori, it is unclear they can generalize well to novel watermark patterns. \textbf{In contrast, our DIP-based evasion operates on a different principle}: it leverages the different rates at which different frequency components are captured during DIP learning through \cref{eq:main-form}, which has been consistently observed in prior DIP literature~\citep{ulyanov2018deep,DBLP_journals_ijcv_ShiMMS22,li2021self,wang2021early}. Specifically, the $G_{\mb \theta}(\mb z)$ term in \cref{eq:main-form} picks up the low-frequency components---which dominate natural images, \textbf{much faster} than picking up the high-frequency components---which tend to be noise-induced, likely watermark-induced for our case. 
\begin{itemize}[nosep,leftmargin=1em]
    \item In \cref{Fig: watermark fourier vis}, we compare the Fourier spectrum of the clean image from \cref{Fig: watermark vis} to those of various watermarked versions: clearly, the spectrum of the clean image concentrates on low frequencies, but different watermark methods reshape the Fourier spectrum by mostly changing the mid-to-high frequencies. For example, RoSteALS and StegaStamp mostly affect the low-mid frequencies, DwtDctSVD and SSL mostly focus on mid-high frequencies, and rivaGAN alters high frequencies. The only exception is TrustMark, whose watermark pattern is hard to observe from the frequency spectrum. 

    \item Next, in \cref{Fig: fourier band error}, we visualize the different learning paces of DIP across different frequency bands, by tracking the frequency band errors (FBEs) similar to \cite{wang2024dmplug,li2023deep,zhuang2024blind}: we first compute the relative per-frequency error in the frequency domain, i.e., $|\mathcal{F}(I_w) - \mathcal{F}(G_{\mb \theta^{(i)}}) / |\mathcal{F} (I_{\mb w})| $, where $\mathcal{F} (\cdot)$ is the discrete Fourier transform, then divide all frequencies into five radial bands from the lowest (1) to the highest (5), and compute the mean errors within each band. It is evident that the lower the frequencies, the faster the FBEs decay. 
    \begin{SCfigure}[][!htbp]
    \includegraphics[width=0.65\textwidth]{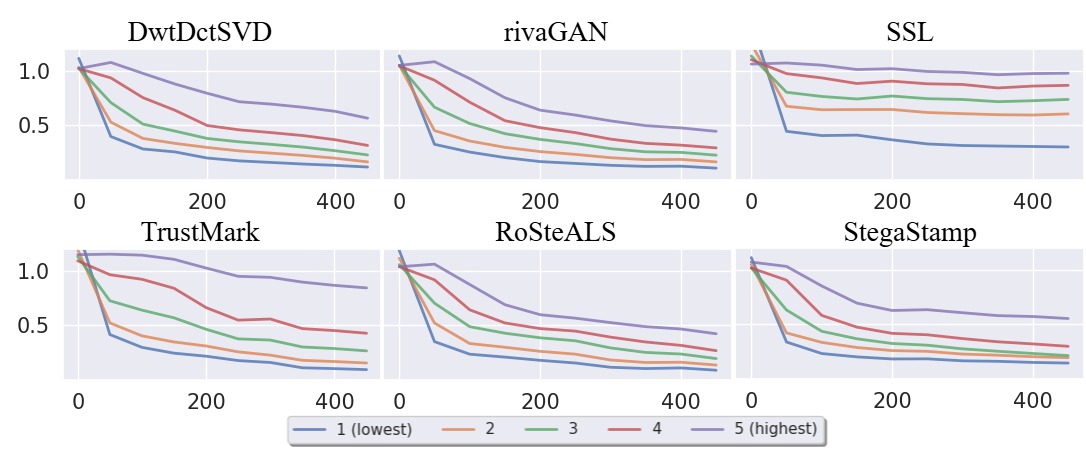}
    \caption{Evolution of the Fourier band errors (FBEs) of DIP's intermediate iterates ($x$-axis: iteration count; $y$-axis: relative band error). We visualize FBEs by dividing all frequency components into five different bands in the radial direction, from the lowest (1) to hightest (5).}
    \label{Fig: fourier band error}
    \vspace{-1em}
\end{SCfigure}
\end{itemize}
Such disparate learning paces imply that certain intermediate iterates enjoy sufficient separation of the low-frequency image content and high-frequency watermark patterns, hence the potential for successful evasion. In summary, compared to alternative purification methods that depend on either explicit noise modeling or data-driven priors, our DIP-based evasion relies on frequency separation that is noise-agnostic and distribution-free. Hence, our DIP-based evasion method largely complements the existing ones. 

\subsection{Remarks on DIP-based watermark evasion}
\label{sec:query-effi}
\cref{alg: DIP evasion} presents a template algorithm of our DIP-based watermark evasion. Depending on the downstream application, we may need to query a decoder (e.g., a detection API) as stated in Line 3 of \cref{alg: DIP evasion}. For example, when the goal is to evade detection while maintaining high image quality, it is essential to query the decoder multiple times to identify the optimal $I_i$. In contrast, for single-shot evasion where image quality is less critical, querying the decoder becomes unnecessary. The consideration applies to all existing black-box evasion methods (e.g., those evaluated in \cref{Sec: exp and results}). 

On the other hand, querying the decoder in every intermediate iterate, as described in \cref{alg: DIP evasion}, may not be possible, especially since the algorithm often requires up to $\sim 1000$ iterations---for example, the decoder server may impose rate limits on individual user to ensure safety and fairness. It is therefore important to reduce the number of decoder queries and improve query efficiency. One potential approach is to selectively query only high-quality iterates---although the clean image $I$ is unknown, the watermarked image $I_{\mb w}$ is, by design (see \cref{eq: watermark def}), very close to $I$, and thus can serve as a proxy for estimating the image quality of intermediate iterates; see \cref{App: psnr-ba} for sample trajectories of image quality and watermark detectability across different watermarks and evasion methods.

Finally, although DIP evasion relies on iterative optimization, it remains computationally efficient in practice. For details on its runtime performance, we refer the reader to \cref{App: runtime comparison}. 

\section{Qualitative and quantitative evaluation}
\label{Sec: exp and results}

\paragraph{Experiment setup}
\textbf{(1) Datasets}: We use images from two large-scale datasets: \textbf{(i)} MS-COCO \citep{lin2014microsoft} composed of 328K real images and \textbf{(ii)} DiffusionDB \citep{wang2022diffusiondb} composed of 14 million high-quality AI-generated images. We randomly sample $2000$ images from each dataset---the typical scale for the robustness evaluation of watermark systems~\citep{an2024benchmarking}, resize them to $512 \times 512$, and generate images with different watermarks, respectively; \textbf{(2) Watermark methods}: We focus on $6$ representative and publicly available \emph{post-processing} watermark methods: DwtDctSVD, rivaGAN, SSL, TrustMark, RoSteALS, and StegaStamp, whose watermark patterns vary in the visibility level and Fourier spectrum; see \cref{Fig: watermark vis,Fig: watermark fourier vis} and \cref{tab: watermark quality intensity} for visual and quantitative comparisons, respectively. We also evaluate on the SOTA \emph{in-processing} TreeRing watermark, where $2000$ watermarked images are generated using \texttt{Gustavosta} stable diffusion prompts from HuggingFace \citep{huggingface2024gustavosta}; \textbf{(3) Evasion methods}: In addition to our DIP-based evasion method described in \cref{alg: DIP evasion}, \footnote{We use the default `skip' network in the original DIP repo: \url{https://github.com/DmitryUlyanov/deep-image-prior}.} we also consider the following classical digital editing methods: \textbf{(i)} brightness, \textbf{(ii)} contrast, \textbf{(iii)} Gaussian noise, \textbf{(iv)} JPEG compression, \textbf{(v)} bm3d denoising, and recent SOTA purification methods: \textbf{(vi)} DiffPure \citep{saberi2023robustness}, \textbf{(vii)} Diffuser (Diffusion attack from \citep{zhao2024invisible}) and \textbf{(viii)} VAE regeneration (VAE attack from \citep{zhao2024invisible}) \footnote{VAE regeneration using model from \cite{cheng2020learned}, which gives the best VAE performance as in \cite{jiang2023evading}.}.

\begin{table*}[!tb]
\caption{Quantitative visual distortion induced by different watermark methods. We report the mean and standard deviation (in parathesis) of all three quality metrics, calculated over $100$ randomly drawn test images. All images are in \texttt{uint8} format with value in $[0, 255]$. RivaGAN has the least visible watermarks, while StegaStamp has the most.}
\label{tab: watermark quality intensity}
\centering
\resizebox{0.9\linewidth}{!}{%
\begin{tabular}{l c ccc c ccc}
\toprule
{ } & & \multicolumn{3}{c}{\textbf{COCO dataset}} & & \multicolumn{3}
{c}{\textbf{DiffusionDB dataset}} \\
\cline{3-5}\cline{7-9}
\vspace{-0.9em}
\\
\small{Watermark} & & \small{PSNR $\uparrow$} & \small{SSIM $\uparrow$} & \small{$90\%$ Quantile $\downarrow$} & & \small{PSNR $\uparrow$} & \small{SSIM $\uparrow$} & \small{$90\%$ Quantile $\downarrow$}\\
\toprule
\vspace{-0.9em}
\\
\small{\textbf{DwtDctSVD}} & & \small{38.31 \small{(2.67)}} & \small{0.98 \small{(0.01)}} & \small{4.66 \small{(0.90)}} & & \small{37.88 \small{(3.60)}} & \small{0.97 \small{(0.02)}} & \small{4.87 \small{(1.56)}} \\
\small{\textbf{rivaGAN}} & & \small{40.57 \small{(0.21)}} & \small{0.98 \small{(0.01)}} & \small{3.00 \small{(0.03)}} & & \small{40.61 \small{(0.24)}} & \small{0.98 \small{(0.01)}} & \small{3.00 \small{(0.03)}} \\
\small{\textbf{SSL}} & & \small{33.03 \small{(0.23)}} & \small{0.88 \small{(0.04)}} & \small{7.19 \small{(0.74)}} & & \small{33.14 \small{(0.55)}} & \small{0.88 \small{(0.04)}} & \small{7.10 \small{(0.72)}} \\
\small{\textbf{TrustMark}} & & \small{40.46 \small{(1.85)}} & \small{0.99 \small{(0.01)}} & \small{3.84 \small{(0.85)}} & & \small{40.67 \small{(2.62)}} & \small{0.99 \small{(0.01)}} & \small{3.84 \small{(1.14)}} \\
\small{\textbf{RoSteALS}} & & \small{30.83 \small{(2.71)}} & \small{0.95 \small{(0.02)}} & \small{12.8 \small{(3.98)}} & & \small{30.88 \small{(4.12)}} & \small{0.95 \small{(0.03)}} & \small{12.5 \small{(5.71)}}\\
\small{\textbf{StegaStamp}} & & \small{28.34 \small{(1.61)}} & \small{0.90 \small{(0.03)}} & \small{15.9 \small{(3.18)}} & & \small{26.58 \small{(2.01)}} & \small{0.84 \small{(0.04)}} & \small{19.3 \small{(4.96)}} \\
\toprule
\end{tabular}}
\vspace{-1em}
\end{table*}%

\paragraph{Evaluation protocol}
As we argue in \cref{Sec: Intro} (see also \cite{an2024benchmarking}), evasion success and image quality are two essential dimensions of watermarking systems. Therefore, we report the \emph{best image quality} each evasion method can achieve while failing watermark detection. To find the ``optimal'' tradeoff image, we perform an exhaustive search over the allowable ranges of the hyperparameters for each evasion method and look for images with \textbf{(i)} watermark undetected and \textbf{(ii)} the highest PSNR value with respect to the watermarked image $I_{\mb w}$; see \cref{sec:query-effi} for justification on why $I_{\mb w}$ is used and \cref{App: hyperparam} for details about all hyperparameters. Finally, to quantify the quality of such images, we use three metrics: \textbf{(i)} PSNR, \textbf{(ii)} Structural Similarity Index Measure (SSIM) \citep{hore2010image} and \textbf{(iii)} $90 \%$ quantile of pixel-wise difference, all with respect to the clean image $I$. The quantile metric mainly serves as a supplement, as PSNR and SSIM focus on the average difference while it reflects the difference on the tail---watermark-induced distortion to an image may be highly localized and hence spatially sparse, which might not be captured by averaging metrics; see \cref{Fig: watermark vis,tab: watermark quality intensity} for a sense of the visual and quantitative distortions caused by different watermark methods. Since the TreeRing watermark lacks a notion of clean image, we use $I_{\mb w}$ as the reference in all evaluation experiments related to TreeRing watermark. 

\begin{table*}[!htbp]
\vspace{-1em}
\caption{The best image quality produced by different evasion methods under different detection threshold $\gamma$ on the COCO dataset. Here, we report the mean value of PSNR-SSIM-$90\%$ Quantile (Q.). We highlight the best evasion method under each watermark and $\gamma$ in \textbf{boldface}. For fair comparison, we mask out cases where one evasion method cannot evade $\ge 90\%$ of the watermarked images.}
\label{tab: watermark evasion diff gamma}
\centering
\resizebox{0.9\linewidth}{!}{%
\begin{tabular}{l c cccc}
\toprule
\multicolumn{2}{l}{~~COCO dataset} & {$\gamma = 0.55$} & {$\gamma = 0.65$} & {$\gamma = 0.75$} & {$\gamma = 0.85$}\\
\toprule
\multicolumn{1}{l}{\textbf{DwtDctSVD}} & \multicolumn{1}{c}{TPR$\uparrow
$ / FPR$\downarrow$} & {1.00 / 0.23} & {0.99 / 0.03} & {0.99 / 0.01} & {0.99 / 0.00} \\
\cdashline{2-6}
\vspace{-0.5em}
\\
{brightness} & {PSNR - SSIM - Q.} & {19.22 - 0.88 - 52.6} & {20.00 - 0.90 - 47.5} & {20.67 - 0.91 - 43.4} & {21.52 - 0.93 - 39.0}\\
{contrast} & & {25.27 - 0.89 - 26.7} & {26.02 - 0.91 - 24.1} & {26.67 - 0.91 - 22.0} & {27.52 - 0.92 - 19.7}\\
{Gaussian noise} & & {16.30 - 0.19 - 66.1} & {17.54 - 0.23 - 56.8} & {18.62 - 0.26 - 50.3} & {20.05 - 0.31 - 42.3}\\
{JPEG} & & {31.52 - 0.88 - 11.2} & {31.74 - 0.89 - 10.9} & {31.89 - 0.89 - 10.7} & {32.13 - 0.90 - 10.4}\\
\rowcolor{Gray}
{bm3d} & & {****} & {****} & {**} & {30.26 - 0.87 - ~9.6~~}\\
\rowcolor{Gray}
{DiffPure} & & {28.20 - 0.79 - 17.3} & {29.61 - 0.83 - 14.2} & {29.90 - 0.84 - 13.6} & {29.94 - 0.84 - 13.6}\\
\rowcolor{Gray}
{Diffuser} & & {**} & {**} & {$^*$25.87 - 0.74 - 20.7~~} & {26.61 - 0.76 - 19.2}\\
\rowcolor{Gray}
{VAE} & & {****} & {$^*$32.38 - 0.88 - 10.5~~} & {33.47 - 0.90 - ~9.2~} & {34.51 - 0.92 - ~8.1~}\\
\cdashline{1-6}
\vspace{-0.95em}
\\
\rowcolor{Gray}
{DIP (ours)} & & \textbf{34.87 - 0.96 - ~7.2~} & \textbf{35.50 - 0.96 - ~6.7~} & \textbf{35.85 - 0.96 - ~6.4~} & \textbf{36.22 - 0.97 - ~6.1~}\\

\toprule
\multicolumn{1}{l}{\textbf{rivaGAN}} & \multicolumn{1}{r}{TPR$\uparrow$ / FPR$\downarrow$} & {0.99 / 0.25} & {0.99 / 0.03} & {0.99 / 0.01} & {0.99 / 0.00} \\
\cdashline{2-6}
\vspace{-0.5em}
\\
{brightness} & {PSNR - SSIM - Q.} & {****} & {~~6.91 - 0.12 - 181~} & {~~7.31 - 0.17 - 173~} & {~8.02 - 0.27 - 159~}\\
{contrast} & {} & {****} & {$^*$13.16 - 0.50 - 90.0~~} & {13.53 - 0.53 - 86.1} & {14.20 - 0.58 - 79.8}\\
{Gaussian noise} & & {11.06 - 0.07 - 121~} & {13.07 - 0.11 - 96.3} & {14.61 - 0.14 - 80.7} & {16.42 - 0.19 - 65.7} \\
{JPEG} & & {****} &
 {**} & {28.22 - 0.79 - 16.7} & {30.16 - 0.85 - 13.2} \\
\rowcolor{Gray}
{bm3d} & & {****} & {****} & {****} & {****}\\
\rowcolor{Gray}
{DiffPure} & & {$^*$26.51 - 0.73 - 21.7} & {28.84 - 0.80 - 15.9} & {29.72 - 0.83 - 14.1} & {29.96 - 0.84 - 13.5}\\
\rowcolor{Gray}
{Diffuser} & & {****} & {$^*$25.40 - 0.72 - 22.1~} & {$^*$26.28 - 0.74 - 20.0~~} & {26.78 - 0.75 - 18.9}\\
\rowcolor{Gray}
{VAE} & & {**} & {$^*$32.34 - 0.88 - \textbf{10.3}~} & {33.28 - 0.90 - ~9.3~} & {34.21 - 0.91 - ~8.3~}\\
\cdashline{1-6}
\vspace{-0.95em}
\\
\rowcolor{Gray}
{DIP (ours)} & & {\textbf{29.87} - \textbf{0.87} - \textbf{17.8}} & {~\textbf{32.64} - \textbf{0.92} - 11.1} & {\textbf{34.02} - \textbf{0.94} - ~\textbf{8.9}~} & {\textbf{35.20} - \textbf{0.95} - ~\textbf{7.4}~}\\

\toprule
\multicolumn{1}{l}{\textbf{SSL}} & \multicolumn{1}{r}{TPR$\uparrow$ / FPR$\downarrow$} & {1.00 / 0.33} & {1.00 / 0.07} & {0.99 / 0.01} & {0.99 / 0.00} \\
\cdashline{2-6}
\vspace{-0.5em}
\\
{brightness} & {PSNR - SSIM - Q.} & {****} & {****} & {****} & {$^*$13.92 - 0.44 - 149~~~~}\\
{contrast} & & {****} & {****} & {**} & {$^*$19.12 - 0.67 - 58.6~~~}\\
{Gaussian noise} & & {$^*$18.89 - 0.28 - 51.4~~} & {21.67 - 0.39 - 36.7} & {23.98 - 0.48 - 27.8} & {25.42 - 0.54 - 22.7}\\
{JPEG} & & {$^*$28.18 - 0.79 - 16.7~~} & {30.05 - 0.84 - 13.4} & {31.28 - 0.86 - 11.4} & {32.73 - 0.89 - ~9.5~~}\\
\rowcolor{Gray}
{bm3d} & & {$^*$27.99 - 0.79 - 14.4~~} & {29.59 - 0.84 - 11.7} & {30.79 - 0.87 - ~\textbf{9.6}~} & {31.18 - 0.89 - ~8.9~~}\\
\rowcolor{Gray}
{DiffPure} & & {27.18 - 0.75 - 19.7} & {28.85 - 0.80 - 15.6} & {29.32 - 0.82 - 14.6} & {29.38 - 0.82 - 14.4}\\
\rowcolor{Gray}
{Diffuser} & & {**} & {$^*$25.39 - 0.68 - 22.1~~} & {$^*$25.80 - 0.69 - 21.1~~} & {$^*$25.90 - 0.69 - 20.8~~~}\\
\rowcolor{Gray}
{VAE} & & {$^*$\textbf{31.21} - \textbf{0.85} - \textbf{11.9}~~} & {$^*$\textbf{32.03} - \textbf{0.88} - \textbf{10.8}~~} & {\textbf{32.83} - \textbf{0.90} - ~9.8~} & {33.50 - 0.91 - ~9.0~}\\
\cdashline{1-6}
\vspace{-0.95em}
\\
\rowcolor{Gray}
{DIP (ours)} & & {23.73 - 0.76 - 31.4} & {28.21 - 0.85 - 17.4} & {31.46 - \textbf{0.90} - 11.2} & {\textbf{33.64} - \textbf{0.92} - ~\textbf{8.4}~}\\

\toprule
\multicolumn{1}{l}{\textbf{TrustMark}} & \multicolumn{1}{r}{TPR$\uparrow$ / FPR$\downarrow$} & {1.00 / 0.13} & {1.00 / 0.00} & {1.00 / 0.00} & {1.00 / 0.00} \\
{brightness} & {PSNR - SSIM - Q.} & {****} & {**} & {~$^*$6.86 - 0.11 - 182~~} & {~$^*$7.27 - 0.17 - 173~~~}\\
{contrast} & & {****} & {**} & {$^*$13.31 - 0.54 - 88.3~~} & {$^*$13.84 - 0.58 - 88.2~~~}\\
{Gaussian noise} & & {~9.57 - 0.05 - 144~~} & {11.17 - 0.07 - 119~~} & {12.72 - 0.09 - 99.0} & {14.49 - 0.13 - 80.9}\\
{JPEG} & & {****} & {****} & {$^*$25.18 - 0.70 - 22.9~~} & {$^*$26.35 - 0.73 - 20.4~~~}\\
\rowcolor{Gray}
{bm3d} & & {------} & {------} & {------} & {------}\\
\rowcolor{Gray}
{DiffPure} & & \textbf{{26.34 - 0.73 - 20.4}} & \textbf{{27.81 - 0.77 - 17.3}} & \textbf{{29.09 - 0.80 - 15.3}} & \textbf{{30.22 - 0.84 - 13.2}}\\
\rowcolor{Gray}
{Diffuser} & & {****} & {$^*$25.62 - 0.74 - 21.1~} & {$^*$26.19 - 0.75 - 19.7~~} & {26.70 - 0.77 - 18.5}\\
\rowcolor{Gray}
{VAE} & & {------} & {------} & {****} & {****}\\
\cdashline{1-6}
\vspace{-0.95em}
\\
\rowcolor{Gray}
{DIP (ours)} & & {**} & {$^*$13.64 - 0.48 - 91.1} & {15.88 - 0.56 - 73.4} & {18.97 - 0.66 - 52.8}\\

\toprule
\multicolumn{1}{l}{\textbf{RoSteALS}} & \multicolumn{1}{c}{TPR$\uparrow
$ / FPR$\downarrow$} & {1.00 / 0.29} & {1.00 / 0.01} & {1.00 / 0.00} & {1.00 / 0.00} \\
\cdashline{2-6}
\vspace{-0.5em}
\\
{brightness} & {PSNR - SSIM - Q.} & {------} & {------} & {------} & {------}\\
{contrast} & & {------} & {------} & {------} & {------}\\
{Gaussian noise} & & {------} & {****} & {****} & {~$^*$9.53 - 0.04 - 145~}\\
{JPEG} & & {------} & {------} & {------} & {****}\\
\rowcolor{Gray}
{bm3d} & & {------} & {------} & {------} & {------}\\
\rowcolor{Gray}
{DiffPure} & & \textbf{$^*$19.12 - 0.49 - 50.6~} & \textbf{22.98 - 0.62 - 30.7} & \textbf{24.91 - 0.69 - 24.3} & \textbf{26.28 - 0.74 - 20.6}\\
\rowcolor{Gray}
{Diffuser} & & {------} & {------} & {------} & {****}\\
\rowcolor{Gray}
{VAE} & & {------} & {------} & {------} & {------}\\
\cdashline{1-6}
\vspace{-0.95em}
\\
\rowcolor{Gray}
{DIP (ours)} & & {****} & {$^*$11.08 - 0.38 - 113~} & {12.09 - 0.42 - 105~} & {16.36 - 0.56 - 73.5} \\

\toprule
\multicolumn{1}{l}{\textbf{StegaStamp}} & \multicolumn{1}{r}{TPR$\uparrow$ / FPR$\downarrow$} & {1.00 / 0.18} & {1.00 / 0.01} & {1.00 / 0.00} & {1.00 / 0.00} \\
{brightness} & {PSNR - SSIM - Q.} & {------} & {------} & {------} & {------}\\
{contrast} & & {------} & {------} & {****} & {$^*$13.22 - 0.52 - 88.9~~~}\\
{Gaussian noise} & & {****} & {~$^*$7.83 - 0.03 - 175~~} & {~9.26 - 0.05 - 148~} & {11.24 - 0.08 - 118~~}\\
{JPEG} & & {****} & {****} & {****} & {**}\\
\rowcolor{Gray}
{bm3d} & & {------} & {------} & {------} & {------}\\
\rowcolor{Gray}
{DiffPure} & & \textbf{{19.57 - 0.50 - 45.6}} & \textbf{{22.42 - 0.61 - 31.5}} & \textbf{{23.81 - 0.66 - 26.7}} & \textbf{{24.67 - 0.70 - 24.1}}\\
\rowcolor{Gray}
{Diffuser} & & {------} & {------} & {****} & {**}\\
\rowcolor{Gray}
{VAE} & & {------} & {------} & {------} & {------}\\
\cdashline{1-6}
\vspace{-0.95em}
\\
\rowcolor{Gray}
{DIP (ours)} & & {**} & {$^*$12.29 - 0.40 - 102~} & {14.00 - 0.47 - 87.5} & {16.05 - 0.55 - 70.9}\\

\toprule
\multicolumn{6}{l}{The following markers are used for the purpose of fair comparison of the best evasion image quality:}\\
\multicolumn{6}{l}{------ Evasion method only successfully evade $<10 \%$ of the watermarked images.}\\
\multicolumn{6}{l}{**** Evasion method only successfully evade $<75\%$ of the watermarked images.}\\
\multicolumn{6}{l}{**~~~~ Evasion method only successfully evade $<90\%$ of the watermarked images.}\\
\multicolumn{6}{l}{*~~~~~~ Evasion method successfully evade $\ge 90\%$ of the watermarked images, but $< 100\%$.}\\
\bottomrule
\end{tabular}}

\vspace{-1em}
\end{table*}%
\begin{table*}[!htbp]
\vspace{-1.5em}
\caption{The best image quality produced by different evasion methods under different detection threshold $\gamma$ on the DiffusionDB dataset. Here, we report the mean value of PSNR-SSIM-$90\%$ Quantile (Q.). We highlight the best evasion method under each watermark and $\gamma$ in \textbf{boldface}. For fair comparison, we mask out cases where one evasion method cannot evade $\ge 90\%$ of the watermarked images.}
\label{tab: watermark evasion diff gamma DiffusionDB}
\centering

\resizebox{0.9\linewidth}{!}{%
\begin{tabular}{l c cccc}
\toprule

\multicolumn{2}{l}{~~DiffusionDB dataset} & {$\gamma = 0.55$} & {$\gamma = 0.65$} & {$\gamma = 0.75$} & {$\gamma = 0.85$}\\
\toprule
\multicolumn{1}{l}{\textbf{DwtDctSVD}} & \multicolumn{1}{c}{TPR$\uparrow
$ / FPR$\downarrow$} & {1.00 / 0.23} & {0.99 / 0.03} & {0.99 / 0.01} & {0.99 / 0.00} \\
\cdashline{2-6}
\vspace{-0.5em}
\\
{brightness} & {PSNR - SSIM - Q.} & {21.25 - 0.90 - 43.0} & {21.94 - 0.91 - 39.1} & {22.54 - 0.92 - 35.9} & {23.34 - 0.93 - 32.3}\\
{contrast} & & {26.89 - 0.91 - 22.2} & {27.56 - 0.92 - 20.1} & {28.11 - 0.93 - 18.6} & {28.85 - 0.94 - 16.8}\\
{Gaussian noise} & & {16.52 - 0.22 - 64.3} & {17.73 - 0.25 - 55.5} & {18.81 - 0.29 - 49.1} & {20.07 - 0.33 - 42.0}\\
{JPEG} & & {30.47 - 0.86 - 13.2} & {30.81 - 0.87 - 12.7} & {31.00 - 0.88 - 12.4} & {31.30 - 0.88 - 12.0}\\
\rowcolor{Gray}
{bm3d} & & {****} & {****} & {**} & {$^*$28.94 - 0.85 - 11.6~~}\\
\rowcolor{Gray}
{DiffPure} & & {27.69 - 0.78 - 18.8} & {29.09 - 0.82 - 15.8} & {29.52 - 0.83 - 15.0} & {29.56 - 0.83 - 14.9}\\
\rowcolor{Gray}
{Diffuser} & & {**} & {**} & {$^*$25.88 - 0.75 - 21.7~~} & {$^*$26.54 - 0.77 - 20.5~~}\\
\rowcolor{Gray}
{VAE} & & {****} & {$^*$31.91 - 0.87 - 11.5~~} & {32.90 - 0.89 - 10.3} & {32.92 - 0.91 - ~9.0~}\\
\cdashline{1-6}
\vspace{-0.95em}
\\
\rowcolor{Gray}
{DIP (ours)} & & \textbf{35.29 - 0.96 - ~6.9~} & \textbf{35.91 - 0.96 - ~6.4~} & \textbf{36.25 - 0.97 - ~6.1~} & \textbf{36.53 - 0.97 - ~5.9~}\\

\toprule
\multicolumn{1}{l}{\textbf{rivaGAN}} & \multicolumn{1}{r}{TPR$\uparrow$ / FPR$\downarrow$} & {0.99 / 0.25} & {0.99 / 0.03} & {0.99 / 0.01} & {0.99 / 0.00} \\
\cdashline{2-6}
\vspace{-0.5em}
\\
{brightness} & {PSNR - SSIM - Q.} & {****} & {$^*$7.38 - 0.13 - 173~} & {~7.83 - 0.19 - 165~} & {~8.70 - 0.31 - 150~}\\
{contrast} & {} & {****} & {$^*$13.47 - 0.51 - 86.8~~} & {13.93 - 0.55 - 82.5} & {14.73 - 0.60 - 75.5}\\
{Gaussian noise} & & {10.96 - 0.08 - 123~} & {12.98 - 0.12 - 97.3} & {14.55 - 0.16 - 81.4} & {16.56 - 0.21 - 64.8} \\
{JPEG} & & {****} &
 {$^*$26.31 - 0.74 - 20.9~} & {$^*$27.84 - 0.79 - 17.5~} & {29.61 - 0.84 - 14.1} \\
\rowcolor{Gray}
{bm3d} & & {****} & {****} & {****} & {****}\\
\rowcolor{Gray}
{DiffPure} & & {26.11 - 0.73 - 22.9} & {28.25 - 0.79 - 16.8} & {28.88 - 0.81 - 15.4} & {29.02 - 0.82 - 15.1}\\
\rowcolor{Gray}
{Diffuser} & & {****} & {$^*$25.43 - 0.73 - 22.5~} & {$^*$26.12 - 0.75 - 20.8~~} & {26.47 - 0.76 - 20.0}\\
\rowcolor{Gray}
{VAE} & & {**} & {$^*$32.29 - 0.88 - \textbf{10.4}~} & {$^*$33.13 - 0.90 - ~9.4~~~} & {33.94 - 0.91 - ~8.5~~}\\
\cdashline{1-6}
\vspace{-0.95em}
\\
\rowcolor{Gray}
{DIP (ours)} & & {\textbf{31.91} - \textbf{0.89} - \textbf{14.5}} & {\textbf{34.48} - \textbf{0.93} - ~9.1~} & {\textbf{35.72} - \textbf{0.95} - ~\textbf{7.4}~} & {\textbf{36.75} - \textbf{0.96} - ~\textbf{6.2}~}\\

\toprule
\multicolumn{1}{l}{\textbf{SSL}} & \multicolumn{1}{r}{TPR$\uparrow$ / FPR$\downarrow$} & {1.00 / 0.33} & {1.00 / 0.07} & {0.99 / 0.01} & {0.99 / 0.00} \\
\cdashline{2-6}
\vspace{-0.5em}
\\
{brightness} & {PSNR - SSIM - Q.} & {****} & {****} & {**} & {$^*$16.91 - 0.53 - 97.9~~~}\\
{contrast} & & {****} & {****} & {**} & {$^*$21.54 - 0.73 - 47.9~~~}\\
{Gaussian noise} & & {$^*$17.43 - 0.26 - 63.8~~} & {20.90 - 0.38 - 41.0} & {23.48 - 0.48 - 29.7} & {25.31 - 0.56 - 23.0}\\
{JPEG} & & {$^*$27.07 - 0.77 - 19.4~~} & {28.97 - 0.82 - 15.4} & {30.47 - 0.85 - 12.7} & {32.39 - 0.89 - ~9.9~~}\\
\rowcolor{Gray}

{bm3d} & & {$^*$26.69 - 0.77 - 16.3~~} & {28.00 - 0.82 - 13.3} & {29.00 - 0.85 - 11.1} & {29.25 - 0.86 - 10.5}\\
\rowcolor{Gray}
{DiffPure} & & {26.67 - 0.75 - 21.0} & {28.05 - 0.79 - 17.3} & {28.40 - 0.80 - 16.3} & {28.43 - 0.80 - 16.2}\\
\rowcolor{Gray}
{Diffuser} & & {**} & {$^*$25.00 - 0.68 - 23.6~~} & {25.43 - 0.69 - 22.5} & {25.49 - 0.70 - 22.4}\\
\rowcolor{Gray}
{VAE} & & {$^*$\textbf{30.26} - \textbf{0.85} - \textbf{12.3}~~} & {$^*$\textbf{31.79} - \textbf{0.88} - \textbf{11.0}~~} & {\textbf{32.43} - \textbf{0.90} - ~\textbf{9.9}~} & {33.01 - 0.91 - ~9.2~}\\
\cdashline{1-6}
\vspace{-0.95em}
\\
\rowcolor{Gray}
{DIP (ours)} & & {23.14 - 0.72 - 35.2} & {27.95 - 0.84 - 18.2} & {31.27 - 0.89 - 11.8} & {\textbf{33.84} - \textbf{0.92} - ~\textbf{8.2}~}\\

\toprule
\multicolumn{1}{l}{\textbf{TrustMark}} & \multicolumn{1}{r}{TPR$\uparrow$ / FPR$\downarrow$} & {1.00 / 0.13} & {1.00 / 0.00} & {1.00 / 0.00} & {1.00 / 0.00} \\
{brightness} & {PSNR - SSIM - Q.} & {****} & {**} & {~$^*$7.09 - 0.10 - 179~~} & {~$^*$7.43 - 0.15 - 172~~~}\\
{contrast} & & {****} & {**} & {$^*$13.52 - 0.55 - 87.3~~} & {$^*$13.98 - 0.59 - 83.2~~~}\\
{Gaussian noise} & & {$^*$9.62 - 0.05 - 143~~~} & {11.17 - 0.08 - 119~~} & {12.67 - 0.10 - 100~} & {14.45 - 0.14 - 81.7}\\
{JPEG} & & {****} & {****} & {$^*$25.06 - 0.69 - 24.5~~} & {$^*$26.08 - 0.72 - 22.4~~~}\\
\rowcolor{Gray}
{bm3d} & & {------} & {------} & {------} & {------}\\
\rowcolor{Gray}
{DiffPure} & & \textbf{{26.60 - 0.73 - 21.8}} & \textbf{{27.91 - 0.77 - 18.7}} & \textbf{{28.99 - 0.79 - 16.6}} & \textbf{{30.07 - 0.83 - 14.4}}\\
\rowcolor{Gray}
{Diffuser} & & {**} & {$^*$25.79 - 0.74 - 21.9~} & {$^*$26.40 - 0.76 - 20.4~~} & {26.96 - 0.77 - 19.2}\\
\rowcolor{Gray}
{VAE} & & {------} & {------} & {****} & {****}\\
\cdashline{1-6}
\vspace{-0.95em}
\\
\rowcolor{Gray}
{DIP (ours)} & & {**} & {$^*$14.16 - 0.5 - 88.4} & {16.19 - 0.57 - 73.0} & {19.07 - 0.65 - 54.6}\\

\toprule
\multicolumn{1}{l}{\textbf{RoSteALS}} & \multicolumn{1}{c}{TPR$\uparrow
$ / FPR$\downarrow$} & {1.00 / 0.29} & {1.00 / 0.01} & {1.00 / 0.00} & {1.00 / 0.00} \\
\cdashline{2-6}
\vspace{-0.5em}
\\
{brightness} & {PSNR - SSIM - Q.} & {------} & {------} & {------} & {------}\\
{contrast} & & {------} & {------} & {------} & {------}\\
{Gaussian noise} & & {------} & {****} & {****} & {~$^*$9.92 - 0.05 - 139~~~}\\
{JPEG} & & {------} & {------} & {------} & {****}\\
\rowcolor{Gray}
{bm3d} & & {------} & {------} & {------} & {------}\\
\rowcolor{Gray}
{DiffPure} & & \textbf{20.86 - 0.57 - 43.7} & \textbf{24.06 - 0.66 - 29.3} & \textbf{25.74 - 0.72 - 23.8} & \textbf{26.93 - 0.76 - 20.6}\\
\rowcolor{Gray}
{Diffuser} & & {------} & {------} & {------} & {****}\\
\rowcolor{Gray}
{VAE} & & {------} & {------} & {------} & {------}\\
\cdashline{1-6}
\vspace{-0.95em}
\\
\rowcolor{Gray}
{DIP (ours)} & & {**} & {$^*$11.67 - 0.42 - 108~~} & {12.83 - 0.46 - 98.9} & {17.47 - 0.60 - 67.0} \\

\toprule
\multicolumn{1}{l}{\textbf{StegaStamp}} & \multicolumn{1}{r}{TPR$\uparrow$ / FPR$\downarrow$} & {1.00 / 0.18} & {1.00 / 0.01} & {1.00 / 0.00} & {1.00 / 0.00} \\
{brightness} & {PSNR - SSIM - Q.} & {------} & {------} & {------} & {------}\\
{contrast} & & {------} & {------} & {****} & {**}\\
{Gaussian noise} & & {****} & {$^*$8.03 - 0.04 - 170~} & {~9.49 - 0.06 - 145~} & {11.66 - 0.09 - 113~}\\
{JPEG} & & {****} & {****} & {****} & {**}\\
\rowcolor{Gray}
{bm3d} & & {------} & {------} & {------} & {------}\\
\rowcolor{Gray}
{DiffPure} & & \textbf{$^*$19.70 - 0.54 - 45.6~} & \textbf{{22.42 - 0.63 - 32.2}} & \textbf{{23.73 - 0.68 - 27.5}} & \textbf{{24.55 - 0.71 - 24.9}}\\
\rowcolor{Gray}
{Diffuser} & & {------} & {------} & {****} & {$^*$21.97 - 0.61 - 32.4~~~}\\
\rowcolor{Gray}
{VAE} & & {------} & {------} & {------} & {------}\\
\cdashline{1-6}
\vspace{-0.95em}
\\
\rowcolor{Gray}
{DIP (ours)} & & {**} & {$^*$13.01 - 0.44 - 96.7~} & {14.69 - 0.50 - 82.7} & {16.54 - 0.57 - 68.5}\\

\toprule
\multicolumn{6}{l}{The following markers are used for the purpose of fair comparison of the best evasion image quality:}\\
\multicolumn{6}{l}{------ Evasion method only successfully evade $<10 \%$ of the watermarked images.}\\
\multicolumn{6}{l}{**** Evasion method only successfully evade $<75\%$ of the watermarked images.}\\
\multicolumn{6}{l}{**~~~~ Evasion method only successfully evade $<90\%$ of the watermarked images.}\\
\multicolumn{6}{l}{*~~~~~~ Evasion method successfully evade $\ge 90\%$ of the watermarked images, but $< 100\%$.}\\
\bottomrule
\end{tabular}}
\vspace{-1em}
\end{table*}%

\begin{table*}[!tb]
\vspace{-1em}
\caption{The best image quality produced by different evasion methods under different detection threshold on Tree-Ring watermarked images. Here, we report the mean value of PSNR-SSIM-$90\%$ Quantile (Q.). We highlight the best number under each watermark and threshold. Note that TreeRing relies on thresholding the $\ell_1$ distance for pattern matching, which is different from $\gamma$ used in other watermarks.  For fair comparison, we mask out cases where one evasion method cannot evade $\ge 90\%$ of the watermarked images.}
\label{tab: tree-ring evasion}
\centering
\resizebox{0.9\linewidth}{!}{%
\begin{tabular}{l c cccc}

\toprule

\textbf{Tree-Ring} & {Threshold} & {$70$} & {$60$} & {$50$} & {$40$}\\
\toprule
{} & \multicolumn{1}{c}{TPR$\uparrow
$ / FPR$\downarrow$} & {1.00 / 0.01} & {1.00 / 0.00} & {0.99 / 0.00} & {0.95 / 0.00} \\
\cdashline{2-6}
\vspace{-0.5em}
\\
{Gaussian noise} & {PSNR - SSIM - Q.} & {**} & {16.49 - 0.19 - 70.6} & {23.06 - 0.42 - 32.9} & {25.75 - 0.53 - 22.3}\\
{JPEG} & & {------} & {$^*$\textbf{28.45} - 0.78 - 17.2~~} & {31.37 - 0.86 - 12.3} & {34.64 - 0.92 - ~8.4~~}\\
\rowcolor{Gray}
{bm3d} & & {****} & {$^*$26.40 - \textbf{0.79} - \textbf{14.6}~~} & {28.66 - 0.85 - 10.2} & {29.47 - 0.88 - ~8.5~~}\\
\rowcolor{Gray}
{DiffPure} & & \textbf{23.20 - 0.63 - 31.8} & {27.79 - 0.77 - 18.7} & {29.79 - 0.83 - 14.1} & {30.16 - 0.84 - 13.0}\\
\rowcolor{Gray}
{Diffuser} & & {------} & {****} & {**} & {$^*$28.01 - 0.84 - 16.1~~}\\
\rowcolor{Gray}
{VAE} & & {------} & {****} & \textbf{$^*$34.06 - 0.89 - ~9.0~~~} & \textbf{$^*$35.58 - 0.93 - ~7.2~~~}\\
\cdashline{1-6}
\vspace{-0.95em}
\\
\rowcolor{Gray}
{DIP (ours)} & & {$^*$12.28 - 0.42 - 102.7} & {17.67 - 0.59 - 64.4} & {26.14 - 0.79 - 26.1} & {34.11 - 0.93 - ~9.6~}\\
\toprule
\multicolumn{6}{l}{The following markers are used for the purpose of fair comparison of the best evasion image quality:}\\
\multicolumn{6}{l}{------ Evasion method only successfully evade $<10 \%$ of the watermarked images.}\\
\multicolumn{6}{l}{**** Evasion method only successfully evade $<75\%$ of the watermarked images.}\\
\multicolumn{6}{l}{**~~~~ Evasion method only successfully evade $<90\%$ of the watermarked images.}\\
\multicolumn{6}{l}{*~~~~~~ Evasion method successfully evade $\ge 90\%$ of the watermarked images, but $< 100\%$.}\\
\bottomrule
\end{tabular}}
\vspace{-1em}
\end{table*}%

\paragraph{Experiment results} 
As mentioned in \cref{Sec: background}, the choice of $\gamma$ in the watermark decoder, as shown in \cref{Eq: detection bar}, is highly task-dependent. It determines the true positive rate (TPR) and the false positive rate (FPR) in watermark detection---TPR measures the fraction of watermarked images correctly detected, and FPR measures the fraction of clean images wrongly flagged as watermarked images. In general, the higher the $\gamma$ used in the decoder, the lower the true positive rate (TPR) and the false positive rate (FPR). A practically useful watermark decoder should have a TPR close to one and a FPR close to zero. On the other hand, the higher the $\gamma$, the higher the quality of the evasion image can achieve. To account for the effect of $\gamma$ in robustness evaluation, we report in \cref{tab: watermark evasion diff gamma} the evasion performance of different methods under different $\gamma$'s on COCO images, and \cref{tab: watermark evasion diff gamma DiffusionDB} on DiffusionDB images, together with the TPR/FPR achieved. 

From \cref{tab: watermark evasion diff gamma,tab: watermark evasion diff gamma DiffusionDB,tab: tree-ring evasion}, we observe that: \textbf{(1)} The effect of $\gamma$ on TPR/FPR values as well as image quality of different watermarks against the detection threshold agrees with our discussion above. This is true also for the TreeRing watermark, whose detection threshold is based on the $\ell_1$ distance, very different from those for other watermark methods; \textbf{(2)} The performance of watermark evasion shown by the column with $\gamma = 0.55$ in \cref{tab: watermark evasion diff gamma,tab: watermark evasion diff gamma DiffusionDB} is not quite meaningful---the very high FPR renders the decoder hardly useful in practice; \textbf{(3)} In other cases, there is no clear winner among all the evasion methods we evaluate. For example, our DIP-based evasion has the best performance on DwtDctSVD and rivaGAN, and is comparable to the best on SSL; DiffPure is the most effective evasion on TrustMark, RoSteALS and StegaStamp; TreeRing seems most vulnerable to JPEG compression. This highlights the need to include diverse sources of evasion methods for faithful robustness evaluation of watermarking systems, as we argue in \cref{sec:why-include-dip}. 

\begin{table}[!htbp]
\centering
\resizebox{0.98\linewidth}{!}{
\begin{tabular}{lcc|ccc|ccc|cc}
\hline
\textbf{Watermarks} & \textbf{(Low) Band 1} & \textbf{Band 2} & \textbf{Band 3} & \textbf{Band 4} & \textbf{Band 5} & \textbf{Band 6} & \textbf{Band 7} & \textbf{Band 8} & \textbf{Band 9} & \textbf{Band 10 (High)} \\
\hline
dwtDctSvd   & 0.025 & 0.044 & 0.059 & 0.074 & 0.091 & 0.103 & 0.109 & 0.109 & 0.110 & 0.371 \\
rivaGAN     & 0.016 & 0.033 & 0.061 & 0.097 & 0.132 & 0.163 & 0.196 &
 0.214 & 0.237 & 0.240 \\
\rowcolor{Gray}
SSL         & 0.011 & 0.036 & 0.070 & 0.119 & 0.174 & 0.222 & 0.289 & 0.331 & 0.377 & 0.765 \\
\rowcolor{Gray}
Trustmark   & 0.017 & 0.075 & \textbf{0.177} & \textbf{0.187} & \textbf{0.189} & 0.175 & 0.148 & 0.113 & 0.094 & 0.109 \\
Rosteals    & 0.085 & 0.282 & \textbf{0.386} & \textbf{0.315} & \textbf{0.411} & 0.436 & 0.475 & 0.556 & 0.620 & 0.307 \\
StegaStamp  & 0.166 & 0.350 & \textbf{0.432} & \textbf{0.443} & \textbf{0.492} & 0.518 & 0.503 & 0.481 & 0.434 & 1.768 \\
\hline
\end{tabular}}
\caption{Average relative Fourier Band Error (FBE) across 10 radial frequency bands for different watermarking methods. Values in bold indicate watermarks with notably high energy in the mid-frequency bands. SSL and Trustmark are also highlighted for comparison: SSL exhibits higher energy in high-frequency components but is vulnerable to DIP evasion, whereas Trustmark shows stronger mid-frequency energy and is more resistant to DIP evasion.}
\label{tab:method_bands}
\end{table}
Moreover, comparing the results of our DIP-based evasion on different watermarks, we observe that evasion images for DwtDctSVD, rivaGAN, and SSL watermarks can achieve very high quality when $\gamma \ge 0.65$, with reasonable TPR/FPR values. This is well expected, as these watermarks mainly cause high-frequency distortions (see \cref{Fig: watermark fourier vis}), and DIP-based evasion is good at separating high- and low-frequency components and thereby largely removing the watermark during iteration, as argued in \cref{sec:why-include-dip}. In contrast,  DIP-based evasion shows limited performance on TrustMark, RoSteALS, and StegaStamp. This is because these watermark methods induce substantial mid- and low-frequency distortions, which DIP picks up in early stages and so are hard to separate from the clean image content. 

To better understand this, we compute the average relative FBE across 10 radial frequency bands using 100 images watermarked by each method. This is done by partitioning the Fourier spectrum of the watermark pattern ($I_w - I$) into different radial bands---a quantitative analysis of qualitative results in \cref{Fig: watermark fourier vis}. As shown in \cref{tab:method_bands}, Trustmark, RoSteALS, and StegaStamp exhibit significantly higher FBE in the lower mid-frequency bands (Bands 3–5) compared to DwtDctSVD, rivaGAN, and SSL. Notably, comparing SSL (whose FBE has larger magnitudes on Band 6-10) and Trustmark (whose FBE has larger magnitudes on Band 3-5), it is even clearer that leveraging lower frequency bands is the key to counter DIP evasions.

\begin{figure*}[!htbp]
\centering
\resizebox{1\linewidth}{!}{%
\includegraphics[width=\textwidth]{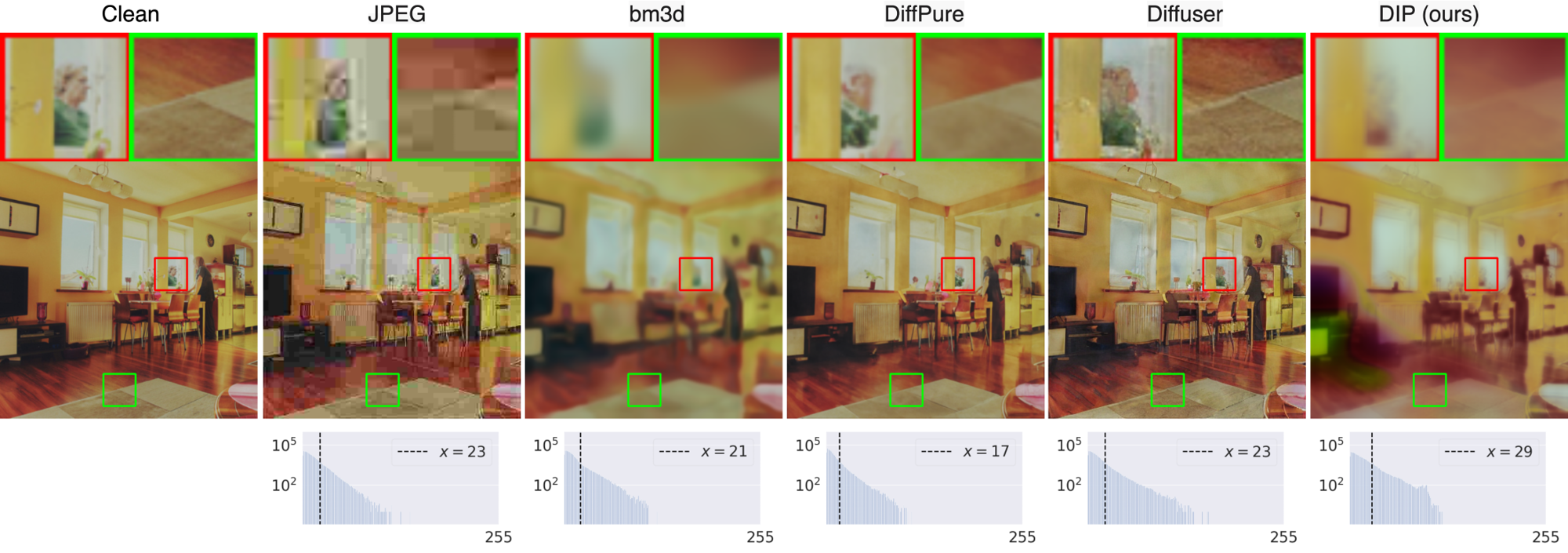}%
}
\vspace{-1em}
\caption{Visualization of the evasion images found by different evasion methods on a StegaStamp watermarked image (with $\gamma = 0.75$; top row) and the respective histograms of the pixel difference ($y-$axis in $\log$ scale) between the evasion image and the clean image (bottom row). The vertical dashed line marks the $90 \%$ quantile.}%
\vspace{-0.5em}
\label{Fig: watermark evasion vis stegaStamp}
\end{figure*}

\section{Conclusion and discussion}
\label{Sec: Summary}

With the results and the analysis above, we can conclude that there is \emph{no universal best evasion methods} for existing watermarking systems. In general, our DIP-based evasion is most effective in evading invisible watermarks that induce high-frequency distortions (e.g., DwtDctSVD, rivaGAN and SSL), and is partially successful  in evading \emph{in-processing} watermarks such as TreeRing. Its limited performance for RoSteALS and StegaStamp implies that exploiting low- and mid-frequency distortions is a viable way for watermarking systems to counteract our DIP-based evasion. Also, for these watermark methods, the regeneration evasion DiffPure has proved effective.  

Moreover, for relatively visible watermarks (e.g., StegaStamp), the evasion images generated by all evasion methods always contain visible artifacts; see \cref{Fig: watermark evasion vis stegaStamp} for an example. Therefore, the future of learning-based watermarks is not all pessimistic: they may not be reliable for copyright protection, but may be promising in misinformation prevention. This is because of the distinct requirements of these two kinds of applications: for copyright protection, watermarks are expected to remain detectable as long as the image content is recognizable even under severe corruptions due to evasion---which may be too hard to achieve. In contrast, to prevent misinformation, it might be sufficient to achieve either of the following to mitigate the harm: \textbf{(i)} the watermark patterns can be detected by eyes, e.g., an overlaid logo or unnatural perturbations such as \cref{Fig: watermark vis intro,Fig: watermark evasion vis stegaStamp}, raising suspicion that the image is already manipulated or fake; \textbf{(ii)} the watermark can be detected by an algorithmic decoder. For this purpose, watermarks such as StegaStamp may be sufficient.

\section{Ethical statement}
\label{Sec: ethical statement}
One potential ethical concern regarding this paper is that it could facilitate the unauthorized removal of watermarks, thereby enabling copyright infringement. However, this concern may be unfounded: Tutorials on using Deep Image Prior (DIP) to remove visible watermarks (e.g., \citet{braindotai2021watermark}) are already publicly available; While research on invisible watermarking is active, such methods have not yet been widely deployed in real-world applications. Rather than promoting misuse, our work aims to proactively identify limitations in current invisible watermarking techniques and to offer concrete recommendations for developing more robust watermarking strategies. 



\bibliography{main}

\begin{thebibliography}{65}
\providecommand{\natexlab}[1]{#1}
\providecommand{\url}[1]{\texttt{#1}}
\expandafter\ifx\csname urlstyle\endcsname\relax
  \providecommand{\doi}[1]{doi: #1}\else
  \providecommand{\doi}{doi: \begingroup \urlstyle{rm}\Url}\fi

\bibitem[An et~al.(2024)An, Ding, Rabbani, Agrawal, Xu, Deng, Zhu, Mohamed, Wen, Goldstein, et~al.]{an2024benchmarking}
Bang An, Mucong Ding, Tahseen Rabbani, Aakriti Agrawal, Yuancheng Xu, Chenghao Deng, Sicheng Zhu, Abdirisak Mohamed, Yuxin Wen, Tom Goldstein, et~al.
\newblock Benchmarking the robustness of image watermarks.
\newblock \emph{arXiv preprint arXiv:2401.08573}, 2024.

\bibitem[Arabi et~al.(2024)Arabi, Feuer, Witter, Hegde, and Cohen]{arabi2024hidden}
Kasra Arabi, Benjamin Feuer, R~Teal Witter, Chinmay Hegde, and Niv Cohen.
\newblock Hidden in the noise: Two-stage robust watermarking for images.
\newblock \emph{arXiv preprint arXiv:2412.04653}, 2024.

\bibitem[Bartz \& Hu(2023)Bartz and Hu]{rueters2023openai}
Diane Bartz and Krystal Hu.
\newblock Openai, google, others pledge to watermark ai content for safety, white house says.
\newblock \emph{https://www.reuters.com/technology/openai-google-others-pledge-watermark-ai-content-safety-white-house-2023-07-21}, 2023.

\bibitem[Bi et~al.(2007)Bi, Sun, Huang, Yang, and Huang]{bi2007robust}
Ning Bi, Qiyu Sun, Daren Huang, Zhihua Yang, and Jiwu Huang.
\newblock Robust image watermarking based on multiband wavelets and empirical mode decomposition.
\newblock \emph{IEEE Transactions on Image Processing}, 16\penalty0 (8):\penalty0 1956--1966, 2007.

\bibitem[Bui et~al.(2023{\natexlab{a}})Bui, Agarwal, and Collomosse]{bui2023trustmark}
Tu~Bui, Shruti Agarwal, and John Collomosse.
\newblock Trustmark: Universal watermarking for arbitrary resolution images.
\newblock \emph{arXiv preprint arXiv:2311.18297}, 2023{\natexlab{a}}.

\bibitem[Bui et~al.(2023{\natexlab{b}})Bui, Agarwal, Yu, and Collomosse]{bui2023rosteals}
Tu~Bui, Shruti Agarwal, Ning Yu, and John Collomosse.
\newblock Rosteals: Robust steganography using autoencoder latent space.
\newblock In \emph{Proceedings of the IEEE/CVF Conference on Computer Vision and Pattern Recognition}, pp.\  933--942, 2023{\natexlab{b}}.

\bibitem[Cheng et~al.(2020)Cheng, Sun, Takeuchi, and Katto]{cheng2020learned}
Zhengxue Cheng, Heming Sun, Masaru Takeuchi, and Jiro Katto.
\newblock Learned image compression with discretized gaussian mixture likelihoods and attention modules.
\newblock In \emph{Proceedings of the IEEE/CVF conference on computer vision and pattern recognition}, pp.\  7939--7948, 2020.

\bibitem[Cox et~al.(1997)Cox, Kilian, Leighton, and Shamoon]{cox1997secure}
Ingemar~J Cox, Joe Kilian, F~Thomson Leighton, and Talal Shamoon.
\newblock Secure spread spectrum watermarking for multimedia.
\newblock \emph{IEEE transactions on image processing}, 6\penalty0 (12):\penalty0 1673--1687, 1997.

\bibitem[Dabov et~al.(2007)Dabov, Foi, Katkovnik, and Egiazarian]{dabov2007image}
Kostadin Dabov, Alessandro Foi, Vladimir Katkovnik, and Karen Egiazarian.
\newblock Image denoising by sparse 3-d transform-domain collaborative filtering.
\newblock \emph{IEEE Transactions on image processing}, 16\penalty0 (8):\penalty0 2080--2095, 2007.

\bibitem[Fan et~al.(2023)Fan, Chen, Wang, and Huang]{fan2023trustworthiness}
Mingyuan Fan, Cen Chen, Chengyu Wang, and Jun Huang.
\newblock On the trustworthiness landscape of state-of-the-art generative models: A comprehensive survey.
\newblock \emph{arXiv preprint arXiv:2307.16680}, 2023.

\bibitem[Fernandez et~al.(2022)Fernandez, Sablayrolles, Furon, J{\'e}gou, and Douze]{fernandez2022watermarking}
Pierre Fernandez, Alexandre Sablayrolles, Teddy Furon, Herv{\'e} J{\'e}gou, and Matthijs Douze.
\newblock Watermarking images in self-supervised latent spaces.
\newblock In \emph{ICASSP 2022-2022 IEEE International Conference on Acoustics, Speech and Signal Processing (ICASSP)}, pp.\  3054--3058. IEEE, 2022.

\bibitem[Fernandez et~al.(2023)Fernandez, Couairon, J{\'e}gou, Douze, and Furon]{fernandez2023stable}
Pierre Fernandez, Guillaume Couairon, Herv{\'e} J{\'e}gou, Matthijs Douze, and Teddy Furon.
\newblock The stable signature: Rooting watermarks in latent diffusion models.
\newblock In \emph{Proceedings of the IEEE/CVF International Conference on Computer Vision}, pp.\  22466--22477, 2023.

\bibitem[Goodfellow et~al.(2014{\natexlab{a}})Goodfellow, Pouget-Abadie, Mirza, Xu, Warde-Farley, Ozair, Courville, and Bengio]{goodfellow2014generative}
Ian Goodfellow, Jean Pouget-Abadie, Mehdi Mirza, Bing Xu, David Warde-Farley, Sherjil Ozair, Aaron Courville, and Yoshua Bengio.
\newblock Generative adversarial nets.
\newblock \emph{Advances in neural information processing systems}, 27, 2014{\natexlab{a}}.

\bibitem[Goodfellow et~al.(2014{\natexlab{b}})Goodfellow, Shlens, and Szegedy]{goodfellow2014explaining}
Ian~J Goodfellow, Jonathon Shlens, and Christian Szegedy.
\newblock Explaining and harnessing adversarial examples.
\newblock \emph{arXiv preprint arXiv:1412.6572}, 2014{\natexlab{b}}.

\bibitem[Gunn et~al.(2024)Gunn, Zhao, and Song]{Gunn2024AnUW}
Sam Gunn, Xuandong Zhao, and Dawn Song.
\newblock An undetectable watermark for generative image models.
\newblock \emph{IACR Cryptol. ePrint Arch.}, 2024:\penalty0 1597, 2024.
\newblock URL \url{https://api.semanticscholar.org/CorpusID:273202887}.

\bibitem[Hore \& Ziou(2010)Hore and Ziou]{hore2010image}
Alain Hore and Djemel Ziou.
\newblock Image quality metrics: Psnr vs. ssim.
\newblock In \emph{2010 20th international conference on pattern recognition}, pp.\  2366--2369. IEEE, 2010.

\bibitem[Hsieh et~al.(2001)Hsieh, Tseng, and Huang]{hsieh2001hiding}
Ming-Shing Hsieh, Din-Chang Tseng, and Yong-Huai Huang.
\newblock Hiding digital watermarks using multiresolution wavelet transform.
\newblock \emph{IEEE Transactions on industrial electronics}, 48\penalty0 (5):\penalty0 875--882, 2001.

\bibitem[Jia et~al.(2021)Jia, Fang, and Zhang]{jia2021mbrs}
Zhaoyang Jia, Han Fang, and Weiming Zhang.
\newblock Mbrs: Enhancing robustness of dnn-based watermarking by mini-batch of real and simulated jpeg compression.
\newblock In \emph{Proceedings of the 29th ACM international conference on multimedia}, pp.\  41--49, 2021.

\bibitem[Jiang et~al.(2023)Jiang, Zhang, and Gong]{jiang2023evading}
Zhengyuan Jiang, Jinghuai Zhang, and Neil~Zhenqiang Gong.
\newblock Evading watermark based detection of ai-generated content.
\newblock In \emph{Proceedings of the 2023 ACM SIGSAC Conference on Computer and Communications Security}, pp.\  1168--1181, 2023.

\bibitem[Jo et~al.(2021)Jo, Chun, and Choi]{DBLP_conf_iccv_JoCC21}
Yeonsik Jo, Se~Young Chun, and Jonghyun Choi.
\newblock Rethinking deep image prior for denoising.
\newblock In \emph{2021 {IEEE/CVF} International Conference on Computer Vision, {ICCV} 2021, Montreal, QC, Canada, October 10-17, 2021}, pp.\  5067--5076. {IEEE}, 2021.
\newblock \doi{10.1109/ICCV48922.2021.00504}.
\newblock URL \url{https://doi.org/10.1109/ICCV48922.2021.00504}.

\bibitem[Kankanhalli et~al.(1999)Kankanhalli, Ramakrishnan, et~al.]{kankanhalli1999adaptive}
Mohan~S Kankanhalli, KR~Ramakrishnan, et~al.
\newblock Adaptive visible watermarking of images.
\newblock In \emph{Proceedings IEEE International Conference on Multimedia Computing and Systems}, volume~1, pp.\  568--573. IEEE, 1999.

\bibitem[Kingma \& Ba(2014)Kingma and Ba]{kingma2014adam}
Diederik~P Kingma and Jimmy Ba.
\newblock Adam: A method for stochastic optimization.
\newblock \emph{arXiv preprint arXiv:1412.6980}, 2014.

\bibitem[Li et~al.(2021)Li, Zhuang, Liang, Peng, Wang, and Sun]{li2021self}
Taihui Li, Zhong Zhuang, Hengyue Liang, Le~Peng, Hengkang Wang, and Ju~Sun.
\newblock Self-validation: Early stopping for single-instance deep generative priors.
\newblock \emph{arXiv preprint arXiv:2110.12271}, 2021.

\bibitem[Li et~al.(2023{\natexlab{a}})Li, Wang, Zhuang, and Sun]{li2023deep}
Taihui Li, Hengkang Wang, Zhong Zhuang, and Ju~Sun.
\newblock Deep random projector: Accelerated deep image prior.
\newblock In \emph{Proceedings of the IEEE/CVF Conference on Computer Vision and Pattern Recognition}, pp.\  18176--18185, 2023{\natexlab{a}}.

\bibitem[Li et~al.(2023{\natexlab{b}})Li, Zhuang, Wang, and Sun]{li_random_2023}
Taihui Li, Zhong Zhuang, Hengkang Wang, and Ju~Sun.
\newblock Random {Projector}: {Efficient} {Deep} {Image} {Prior}.
\newblock In \emph{{ICASSP} 2023 - 2023 {IEEE} {International} {Conference} on {Acoustics}, {Speech} and {Signal} {Processing} ({ICASSP})}, pp.\  1--5, June 2023{\natexlab{b}}.
\newblock \doi{10.1109/ICASSP49357.2023.10097088}.

\bibitem[Li et~al.(2024)Li, Lahiri, Dai, and Mayer]{jdd_doubledip}
Taihui Li, Anish Lahiri, Yutong Dai, and Owen Mayer.
\newblock Joint demosaicing and denoising with double deep image priors.
\newblock In \emph{ICASSP 2024 - 2024 IEEE International Conference on Acoustics, Speech and Signal Processing (ICASSP)}, pp.\  4005--4009, 2024.
\newblock \doi{10.1109/ICASSP48485.2024.10448384}.

\bibitem[Lin et~al.(2014)Lin, Maire, Belongie, Hays, Perona, Ramanan, Doll{\'a}r, and Zitnick]{lin2014microsoft}
Tsung-Yi Lin, Michael Maire, Serge Belongie, James Hays, Pietro Perona, Deva Ramanan, Piotr Doll{\'a}r, and C~Lawrence Zitnick.
\newblock Microsoft coco: Common objects in context.
\newblock In \emph{Computer Vision--ECCV 2014: 13th European Conference, Zurich, Switzerland, September 6-12, 2014, Proceedings, Part V 13}, pp.\  740--755. Springer, 2014.

\bibitem[Luo et~al.(2020)Luo, Zhan, Chang, Yang, and Milanfar]{luo2020distortion}
Xiyang Luo, Ruohan Zhan, Huiwen Chang, Feng Yang, and Peyman Milanfar.
\newblock Distortion agnostic deep watermarking.
\newblock In \emph{Proceedings of the IEEE/CVF conference on computer vision and pattern recognition}, pp.\  13548--13557, 2020.

\bibitem[Mataev et~al.(2019)Mataev, Milanfar, and Elad]{mataev2019deepred}
Gary Mataev, Peyman Milanfar, and Michael Elad.
\newblock Deepred: Deep image prior powered by red.
\newblock In \emph{Proceedings of the IEEE/CVF International Conference on Computer Vision Workshops}, pp.\  0--0, 2019.

\bibitem[Mishra(2022)]{remove2022remove}
Jennifer Mishra.
\newblock How to remove dall-e watermark, 2022.
\newblock URL \url{https://www.youtube.com/watch?v=6EMROCxGCIA}.

\bibitem[Morkel et~al.(2005)Morkel, Eloff, and Olivier]{morkel2005overview}
Tayana Morkel, Jan~HP Eloff, and Martin~S Olivier.
\newblock An overview of image steganography.
\newblock In \emph{ISSA}, volume~1, pp.\  1--11, 2005.

\bibitem[Mumuni \& Mumuni(2022)Mumuni and Mumuni]{mumuni2022data}
Alhassan Mumuni and Fuseini Mumuni.
\newblock Data augmentation: A comprehensive survey of modern approaches.
\newblock \emph{Array}, 16:\penalty0 100258, 2022.

\bibitem[Navas et~al.(2008)Navas, Ajay, Lekshmi, Archana, and Sasikumar]{navas2008dwt}
KA~Navas, Mathews~Cheriyan Ajay, M~Lekshmi, Tampy~S Archana, and M~Sasikumar.
\newblock Dwt-dct-svd based watermarking.
\newblock In \emph{2008 3rd international conference on communication systems software and middleware and workshops (COMSWARE'08)}, pp.\  271--274. IEEE, 2008.

\bibitem[Pereira \& Pun(1999)Pereira and Pun]{pereira2000fast}
Shelby Pereira and Thierry Pun.
\newblock Fast robust template matching for affine resistant image watermarks.
\newblock In \emph{International Workshop on Information Hiding}, pp.\  199--210. Springer, 1999.

\bibitem[Pereira \& Pun(2000)Pereira and Pun]{pereira2000robust}
Shelby Pereira and Thierry Pun.
\newblock Robust template matching for affine resistant image watermarks.
\newblock \emph{IEEE transactions on image Processing}, 9\penalty0 (6):\penalty0 1123--1129, 2000.

\bibitem[Qayyum et~al.(2023)Qayyum, Ilahi, Shamshad, Boussa{\"{\i}}d, Bennamoun, and Qadir]{DBLP_journals_pami_QayyumISBBQ23}
Adnan Qayyum, Inaam Ilahi, Fahad Shamshad, Farid Boussa{\"{\i}}d, Mohammed Bennamoun, and Junaid Qadir.
\newblock Untrained neural network priors for inverse imaging problems: {A} survey.
\newblock \emph{{IEEE} Trans. Pattern Anal. Mach. Intell.}, 45\penalty0 (5):\penalty0 6511--6536, 2023.
\newblock \doi{10.1109/TPAMI.2022.3204527}.
\newblock URL \url{https://doi.org/10.1109/TPAMI.2022.3204527}.

\bibitem[Ramesh et~al.(2022)Ramesh, Dhariwal, Nichol, Chu, and Chen]{ramesh2022hierarchical}
Aditya Ramesh, Prafulla Dhariwal, Alex Nichol, Casey Chu, and Mark Chen.
\newblock Hierarchical text-conditional image generation with clip latents.
\newblock \emph{arXiv preprint arXiv:2204.06125}, 1\penalty0 (2):\penalty0 3, 2022.

\bibitem[Rishik(2020)]{braindotai2021watermark}
Mouryam Rishik.
\newblock Watermark removal using deep image priors with pytorch.
\newblock \emph{https://github.com/braindotai/Watermark-Removal-Pytorch/commits/master}, 2020.

\bibitem[Rombach et~al.(2022)Rombach, Blattmann, Lorenz, Esser, and Ommer]{rombach2022high}
Robin Rombach, Andreas Blattmann, Dominik Lorenz, Patrick Esser, and Bj{\"o}rn Ommer.
\newblock High-resolution image synthesis with latent diffusion models.
\newblock In \emph{Proceedings of the IEEE/CVF conference on computer vision and pattern recognition}, pp.\  10684--10695, 2022.

\bibitem[Saberi et~al.(2023)Saberi, Sadasivan, Rezaei, Kumar, Chegini, Wang, and Feizi]{saberi2023robustness}
Mehrdad Saberi, Vinu~Sankar Sadasivan, Keivan Rezaei, Aounon Kumar, Atoosa Chegini, Wenxiao Wang, and Soheil Feizi.
\newblock Robustness of ai-image detectors: Fundamental limits and practical attacks.
\newblock \emph{arXiv preprint arXiv:2310.00076}, 2023.

\bibitem[Santana()]{huggingface2024gustavosta}
Gustavo Santana.
\newblock Stable diffusion dataset.
\newblock \url{https://huggingface.co/datasets/Gustavosta/Stable-Diffusion-Prompts}.
\newblock Accessed: 2025-02-16.

\bibitem[Shi et~al.(2022{\natexlab{a}})Shi, Mettes, Maji, and Snoek]{DBLP_journals_ijcv_ShiMMS22}
Zenglin Shi, Pascal Mettes, Subhransu Maji, and Cees G.~M. Snoek.
\newblock On measuring and controlling the spectral bias of the deep image prior.
\newblock \emph{Int. J. Comput. Vis.}, 130\penalty0 (4):\penalty0 885--908, 2022{\natexlab{a}}.
\newblock \doi{10.1007/S11263-021-01572-7}.
\newblock URL \url{https://doi.org/10.1007/s11263-021-01572-7}.

\bibitem[Shi et~al.(2022{\natexlab{b}})Shi, Mettes, Maji, and Snoek]{shi2022measuring}
Zenglin Shi, Pascal Mettes, Subhransu Maji, and Cees~GM Snoek.
\newblock On measuring and controlling the spectral bias of the deep image prior.
\newblock \emph{International Journal of Computer Vision}, 130\penalty0 (4):\penalty0 885--908, 2022{\natexlab{b}}.

\bibitem[Tancik et~al.(2020)Tancik, Mildenhall, and Ng]{tancik2020stegastamp}
Matthew Tancik, Ben Mildenhall, and Ren Ng.
\newblock Stegastamp: Invisible hyperlinks in physical photographs.
\newblock In \emph{Proceedings of the IEEE/CVF conference on computer vision and pattern recognition}, pp.\  2117--2126, 2020.

\bibitem[Tirer et~al.(2023)Tirer, Se, Chun, Eldar, {\textcopyright}SHUTTERSTOCK.COM, and Krasovitckii]{Tirer2023DeepIL}
Tom Tirer, Raja~Giryes Se, Young Chun, Yonina~C. Eldar, {\textcopyright}SHUTTERSTOCK.COM, and Andrew Krasovitckii.
\newblock Deep internal learning: Deep learning from a single input.
\newblock \emph{IEEE Signal Processing Magazine}, 41:\penalty0 40--57, 2023.
\newblock URL \url{https://api.semanticscholar.org/CorpusID:266174018}.

\bibitem[Tirkel et~al.(1993)Tirkel, Rankin, Van~Schyndel, Ho, Mee, and Osborne]{tirkel1993electronic}
Anatol~Z Tirkel, GA~Rankin, RM~Van~Schyndel, WJ~Ho, NRA Mee, and Charles~F Osborne.
\newblock Electronic watermark.
\newblock \emph{Digital Image Computing, Technology and Applications (DICTA’93)}, pp.\  666--673, 1993.

\bibitem[Ulyanov et~al.(2018)Ulyanov, Vedaldi, and Lempitsky]{ulyanov2018deep}
Dmitry Ulyanov, Andrea Vedaldi, and Victor Lempitsky.
\newblock Deep image prior.
\newblock In \emph{Proceedings of the IEEE conference on computer vision and pattern recognition}, pp.\  9446--9454, 2018.

\bibitem[Voyatzis \& Pitas(1999)Voyatzis and Pitas]{voyatzis1999protecting}
George Voyatzis and Ioannis Pitas.
\newblock Protecting digital image copyrights: a framework.
\newblock \emph{IEEE Computer Graphics and Applications}, 19\penalty0 (1):\penalty0 18--24, 1999.

\bibitem[Wang et~al.(2021)Wang, Li, Zhuang, Chen, Liang, and Sun]{wang2021early}
Hengkang Wang, Taihui Li, Zhong Zhuang, Tiancong Chen, Hengyue Liang, and Ju~Sun.
\newblock Early stopping for deep image prior.
\newblock \emph{arXiv preprint arXiv:2112.06074}, 2021.

\bibitem[Wang et~al.(2024)Wang, Zhang, Li, Wan, Chen, and Sun]{wang2024dmplug}
Hengkang Wang, Xu~Zhang, Taihui Li, Yuxiang Wan, Tiancong Chen, and Ju~Sun.
\newblock Dmplug: A plug-in method for solving inverse problems with diffusion models.
\newblock \emph{arXiv preprint arXiv:2405.16749}, 2024.

\bibitem[Wang et~al.(2022)Wang, Montoya, Munechika, Yang, Hoover, and Chau]{wang2022diffusiondb}
Zijie~J Wang, Evan Montoya, David Munechika, Haoyang Yang, Benjamin Hoover, and Duen~Horng Chau.
\newblock Diffusiondb: A large-scale prompt gallery dataset for text-to-image generative models.
\newblock \emph{arXiv preprint arXiv:2210.14896}, 2022.

\bibitem[Wen \& Aydore(2019)Wen and Aydore]{wen2019romark}
Bingyang Wen and Sergul Aydore.
\newblock Romark: A robust watermarking system using adversarial training.
\newblock \emph{arXiv preprint arXiv:1910.01221}, 2019.

\bibitem[Wen et~al.(2023)Wen, Kirchenbauer, Geiping, and Goldstein]{wen2023tree}
Yuxin Wen, John Kirchenbauer, Jonas Geiping, and Tom Goldstein.
\newblock Tree-ring watermarks: Fingerprints for diffusion images that are invisible and robust.
\newblock \emph{arXiv preprint arXiv:2305.20030}, 2023.

\bibitem[Wendling()]{bbc2024ai}
Mike Wendling.
\newblock Ai can be easily used to make fake election photos - report.
\newblock \emph{https://www.bbc.com/news/world-us-canada-68471253}.

\bibitem[Yang et~al.(2024)Yang, Zeng, Chen, Fang, Zhang, and Yu]{Yang2024GaussianSP}
Zijin Yang, Kai Zeng, Kejiang Chen, Han Fang, Wei~Ming Zhang, and Neng~H. Yu.
\newblock Gaussian shading: Provable performance-lossless image watermarking for diffusion models.
\newblock \emph{2024 IEEE/CVF Conference on Computer Vision and Pattern Recognition (CVPR)}, pp.\  12162--12171, 2024.
\newblock URL \url{https://api.semanticscholar.org/CorpusID:269004589}.

\bibitem[Yu et~al.(2021)Yu, Skripniuk, Abdelnabi, and Fritz]{yu2021artificial}
Ning Yu, Vladislav Skripniuk, Sahar Abdelnabi, and Mario Fritz.
\newblock Artificial fingerprinting for generative models: Rooting deepfake attribution in training data.
\newblock In \emph{Proceedings of the IEEE/CVF International conference on computer vision}, pp.\  14448--14457, 2021.

\bibitem[Zhang et~al.(2019{\natexlab{a}})Zhang, Cuesta-Infante, Xu, and Veeramachaneni]{zhang2019steganogan}
Kevin~Alex Zhang, Alfredo Cuesta-Infante, Lei Xu, and Kalyan Veeramachaneni.
\newblock Steganogan: High capacity image steganography with gans.
\newblock \emph{arXiv preprint arXiv:1901.03892}, 2019{\natexlab{a}}.

\bibitem[Zhang et~al.(2019{\natexlab{b}})Zhang, Xu, Cuesta-Infante, and Veeramachaneni]{zhang2019robust}
Kevin~Alex Zhang, Lei Xu, Alfredo Cuesta-Infante, and Kalyan Veeramachaneni.
\newblock Robust invisible video watermarking with attention.
\newblock \emph{arXiv preprint arXiv:1909.01285}, 2019{\natexlab{b}}.

\bibitem[Zhang et~al.(2018)Zhang, Isola, Efros, Shechtman, and Wang]{zhang2018unreasonable}
Richard Zhang, Phillip Isola, Alexei~A Efros, Eli Shechtman, and Oliver Wang.
\newblock The unreasonable effectiveness of deep features as a perceptual metric.
\newblock In \emph{Proceedings of the IEEE conference on computer vision and pattern recognition}, pp.\  586--595, 2018.

\bibitem[Zhao et~al.(2024{\natexlab{a}})Zhao, Gunn, Christ, Fairoze, Fabrega, Carlini, Garg, Hong, Nasr, Tram{\`e}r, Jha, Li, Wang, and Song]{Zhao2024SoKWF}
Xuandong Zhao, Sam Gunn, Miranda Christ, Jaiden Fairoze, Andres Fabrega, Nicholas Carlini, Sanjam Garg, Sanghyun Hong, Milad Nasr, Florian Tram{\`e}r, Somesh Jha, Lei Li, Yu-Xiang Wang, and Dawn Song.
\newblock Sok: Watermarking for ai-generated content.
\newblock \emph{ArXiv}, abs/2411.18479, 2024{\natexlab{a}}.
\newblock URL \url{https://api.semanticscholar.org/CorpusID:274305578}.

\bibitem[Zhao et~al.(2024{\natexlab{b}})Zhao, Zhang, Su, Vasan, Grishchenko, Kruegel, Vigna, Wang, and Li]{zhao2024invisible}
Xuandong Zhao, Kexun Zhang, Zihao Su, Saastha Vasan, Ilya Grishchenko, Christopher Kruegel, Giovanni Vigna, Yu-Xiang Wang, and Lei Li.
\newblock Invisible image watermarks are provably removable using generative ai.
\newblock \emph{Advances in neural information processing systems}, 37:\penalty0 8643--8672, 2024{\natexlab{b}}.

\bibitem[Zhu et~al.(2018)Zhu, Kaplan, Johnson, and Fei-Fei]{zhu2018hidden}
Jiren Zhu, Russell Kaplan, Justin Johnson, and Li~Fei-Fei.
\newblock Hidden: Hiding data with deep networks.
\newblock In \emph{Proceedings of the European conference on computer vision (ECCV)}, pp.\  657--672, 2018.

\bibitem[Zhuang(2023)]{zhuang2023advancing}
Zhong Zhuang.
\newblock \emph{Advancing Deep Learning for Scientific Inverse Problems}.
\newblock PhD thesis, University of Minnesota, 2023.

\bibitem[Zhuang et~al.(2023)Zhuang, Yang, Hofmann, Barmherzig, Sun, Yang, Hofmann, Barmherzig, and Sun]{zhuang_practical_2023}
Zhong Zhuang, David Yang, Felix Hofmann, David Barmherzig, Ju~Sun, David Yang, Felix Hofmann, David Barmherzig, and Ju~Sun.
\newblock Practical phase retrieval using double deep image priors.
\newblock \emph{Electronic Imaging}, 35:\penalty0 1--6, January 2023.
\newblock ISSN 2470-1173.
\newblock \doi{10.2352/EI.2023.35.14.COIMG-153}.
\newblock URL \url{https://library.imaging.org/ei/articles/35/14/COIMG-153}.
\newblock Publisher: Society for Imaging Science and Technology.

\bibitem[Zhuang et~al.(2024)Zhuang, Li, Wang, and Sun]{zhuang2024blind}
Zhong Zhuang, Taihui Li, Hengkang Wang, and Ju~Sun.
\newblock Blind image deblurring with unknown kernel size and substantial noise.
\newblock \emph{International Journal of Computer Vision}, 132\penalty0 (2):\penalty0 319--348, 2024.

\end{thebibliography}
\bibliographystyle{tmlr}

\appendix
\section{DALLE-2 visible watermark}
\label{App: DALLE-2 Example}
\begin{SCfigure}[][!htbp]
\centering
\includegraphics[width=0.4\textwidth]{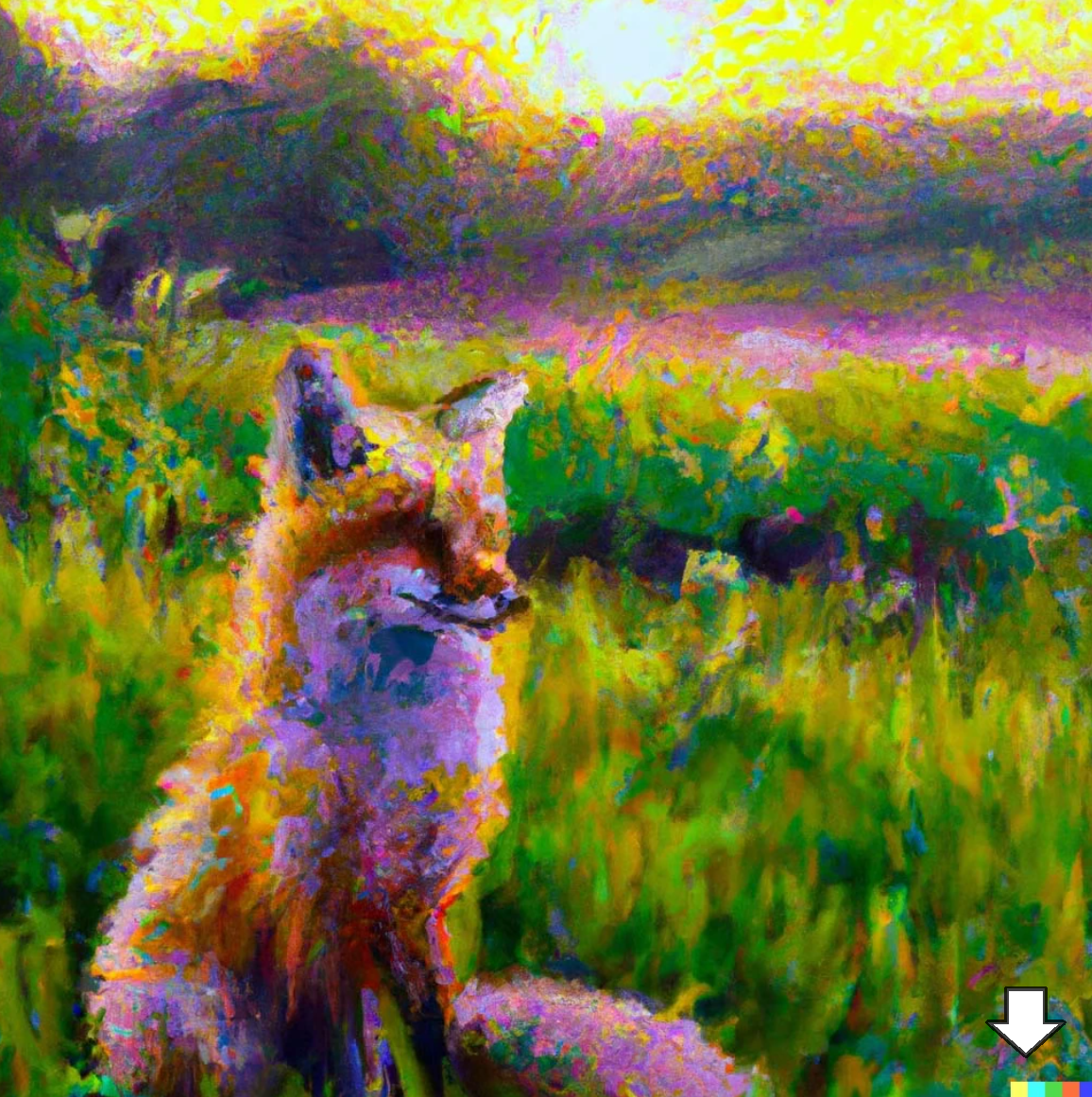}%
\caption{An example image generated by \texttt{DALLE-2} from the official website: \url{https://openai.com/index/dall-e-2/}, where the color code watermark is visible at the bottom-right corner.}%
\vspace{-0.5em}
\label{Fig App: DALLE 2 Vis}
\end{SCfigure}

An example of a visible \texttt{DALLE-2} watermark in colored blocks is shown \cref{Fig App: DALLE 2 Vis}.

\section{Ranges of hyperparameters of various evasion methods}
\label{App: hyperparam}
\begin{table*}[!htbp]
\caption{Details of hyperparameters used in our exhaustive search}
\label{App tab: evasion hyperparam}
\centering
\resizebox{0.9\linewidth}{!}{%
\begin{tabular}{l c cccc}
\toprule
\small{Evasion method} & & \small{Hyperparameter} & \small{Search range} & \small{Search resolution} & \small{Common default value}\\
\toprule
\vspace{-0.9em}
\\
\small{\textbf{brightness}} & & \small{Enhancement factor} & \small{[0.01,  1]} & \small{0.01} & \small{0.5} \\
\small{\textbf{contrast}} & & \small{Enhancement factor} & \small{[0.01,  1]} & \small{0.01} & \small{0.5}\\
\small{\textbf{Gaussian Noise}} & & \small{Standard deviation} & \small{[0.01,  1]} & \small{0.01} & \small{0.1} \\
\small{\textbf{JPEG}} & & \small{Quality factor} & \small{[1,  100]} & \small{1} & \small{50} \\
\small{\textbf{bm3d}} & & \small{Noise standard deviation} & \small{[0.1, 5]} & \small{0.05} & \small{0.1} \\
\small{\textbf{DiffPure}} & & \small{Diffusion noise level} & \small{[0.1, 1]} & \small{0.1} & \small{0.1 - 0.3} \\
\small{\textbf{Diffuser}} & & \small{Diffusion inverse steps} & \small{[10, 100]} & \small{10} & \small{60} \\
\small{\textbf{VAE-Cheng2020}} & & \small{VAE compression quality indenx} & \small{[1, 6]} & \small{1} & \small{3} \\
\small{\textbf{DIP} (ours)} & & \small{Number of iteration} & \small{[1, 500]} & \small{10} & \small{-} \\
\toprule
\end{tabular}}
\vspace{-1em}
\end{table*}%
\cref{App tab: evasion hyperparam} presents the range of each hyperparameter and the grid resolution in our exhaustive search. If we simply use these typical values without exhaustive search, it is very likely that we will overestimate the level of robustness and underestimate the evasion image quality. \cref{Fig: DIP evade algo 2} shows an example of the exhaustive search process on our DIP-based evasion.

In \cref{App tab: evasion hyperparam}, the ``quality index'' for the VAE refers to the selection of a specific pretrained model (among the six provided in the original work), hence the minimal grid resolution for this hyperparameter is 1. The final column lists the default values commonly used in the literature for robustness evaluation (e.g., \cite{zhao2024invisible, saberi2023robustness}). To the best of our knowledge, the rationale behind the choice of these default settings is not well documented.

\section{Runtime comparison of different watermark evasions}
\label{App: runtime comparison}
\begin{table}[!htbp]
\centering
\caption{Runtime comparison of evading a single watermarked image by different watermark evasion methods}
\label{app tab:runtime comparison}
\begin{tabular}{l c c}
\toprule
Method & Evasion configuration & Time (s) \\
\midrule
Diffuser & 10 different diffusion steps & 3.85 \\
BM3D & 5 different standard deviation parameters & 18.2 \\
DIP & 500 iterations & 26.6 \\
DiffPure & 10 different diffusion steps & 290.75\\
\bottomrule
\end{tabular}
\end{table}
We perform a runtime comparison of different watermark evasion methods mentioned in this paper. The experiment is carried out on a local desktop with \texttt{Windows OS} equipped with an \texttt{Intel Core i7-12700K} processor and an \texttt{NVIDIA RTX 3080 GPU}, on the evasions of a single image. The result can be found in \cref{app tab:runtime comparison}: Our DIP-based evasion is comparable to Diffuser and BM3D, and is substantially faster than DiffPure. Moreover, we note that recent research has tried to accelerate DIP-based denoising, e.g., \cite{li2023deep}, which---although beyond the scope of this paper---represents promising directions for further improving the computational efficiency of our DIP-based watermark evasion. 

\section{Image quality v.s. evasion success} 
\label{App: psnr-ba}


\begin{figure}[!htbp]
\centering
\includegraphics[width=1\textwidth]{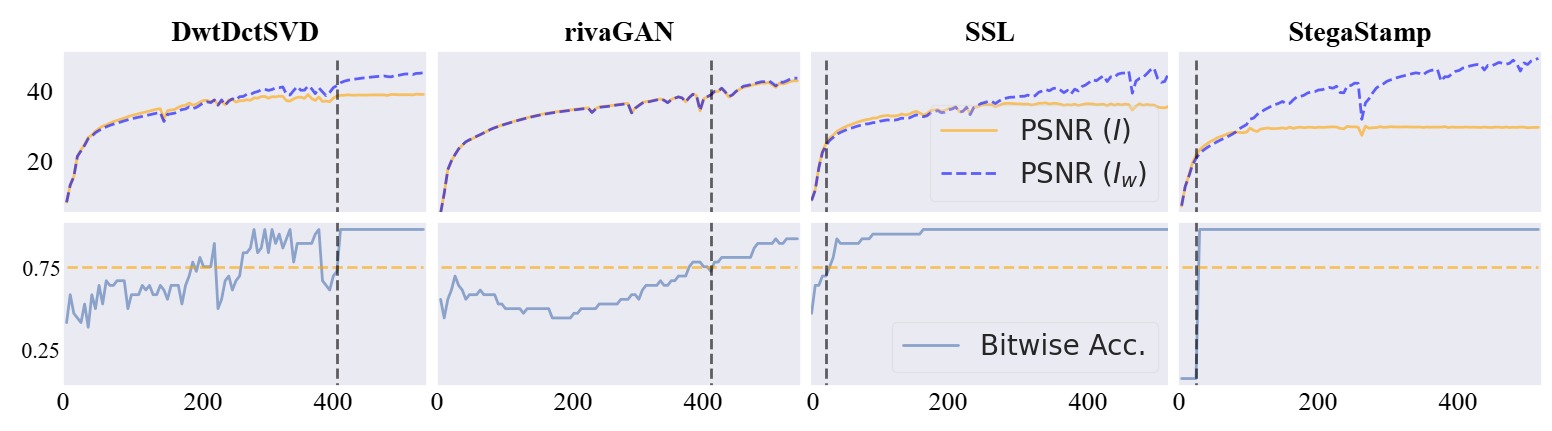}
\vspace{-1.5em}
\caption{The PSNR trajectories of DIP-based evasion (top row) with respect to the watermarked image (in blue) and with respect to the original image (orange), and the corresponding trajectory of evasion performance (bottom row, measured by $BA$), for different watermark systems. Consider the BA threshold $\gamma=0.75$ for detection (marked by horizontal orange lines). The vertical black lines mark the best-quality evasion images.}%
\vspace{-1em}
\label{Fig: DIP evade algo 2}
\end{figure}
\cref{Fig: DIP evade algo 2} presents several illustrative trajectories of the relationship between image quality and the evasion success evaluated on our DIP-based evasion.

\section{Additional visualization of evasion image quality on rivaGAN}
\label{Sec App: rivaGan evasion patterns}
\begin{figure*}[!htbp]
\centering
\resizebox{0.9\linewidth}{!}{%
\includegraphics[width=\textwidth]{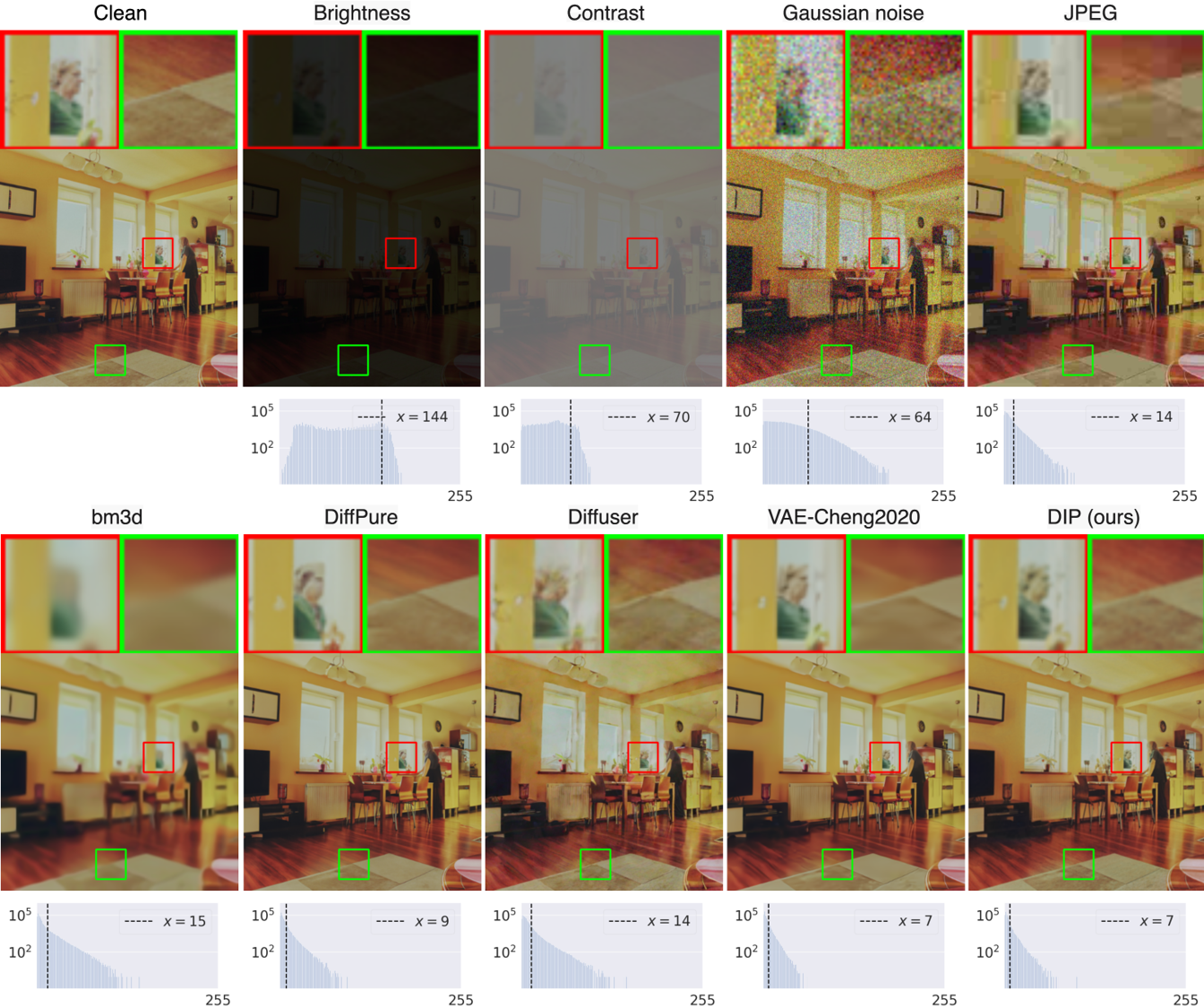}%
}
\vspace{-1em}
\caption{Visualization of the evasion images found by different evasion methods on a rivaGAN watermarked image (with $\gamma = 0.75$; top row) and the respective histograms of the pixel difference ($y-$axis in $\log$ scale) between the evasion image and the clean image (bottom row). The vertical dashed line marks the $90 \%$ quantile. We can observe that the evasion image produced by DIP has almost no loss of image quality.}%
\vspace{-0.5em}
\label{Fig: watermark evasion vis rivaGAN}
\end{figure*}

\cref{Fig: watermark evasion vis rivaGAN} presents evasion images with the best visual quality from the different evasion methods considered in this paper. Note that all other methods except for DIP results in visual artifacts in the evasion image, e.g., pixel jitters (Gaussian Noise, JPEG and Diffuser) or overly smoothing effect (bm3d, DiffPure and VAE).

\section{Additional discussion on the WIND watermark}
\citet{arabi2024hidden} proposes an inpainting-based variant of the WIND watermark ($\text{WIND}_{\text{inpainting}}$), whose underlying principle is similar to that of TreeRing~\citep{wen2023tree} and aims to extend diffusion-based in-processing watermarking methods to post-processing scenarios. The primary focus of \citet{arabi2024hidden} is the identification (i.e., recovery) of the watermark key, but not watermark detection---the binary classification of watermarked vs. non-watermarked images; see the evaluation in Table 1 in \citet{wen2023tree}. 

\begin{figure}[!htbp] 
\centering
\includegraphics[width=0.5\textwidth]{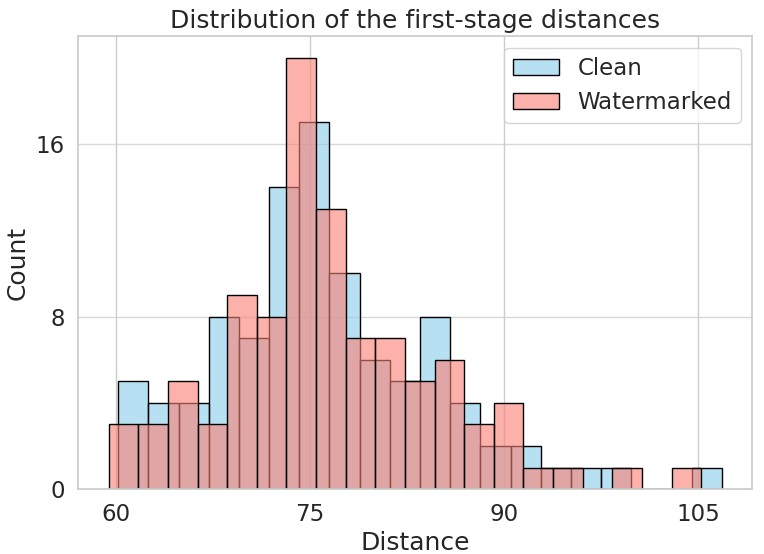}%
\caption{Distribution of the first-stage distance for clean and watermarked images decoded by $\text{WIND}_{\text{inpainting}}$}%
\vspace{-0.5em}
\label{Fig App: First-stage distance}
\end{figure}

To evaluate the detection performance of $\text{WIND}_{\text{inpainting}}$, we perform the following experiment: We sample 100 images from the COCO dataset and generate the corresponding watermarked versions using $\text{WIND}_{\text{inpainting}}$ \citep{arabi2024hidden}. We then apply the $\text{WIND}_{\text{inpainting}}$ decoder to all 200 images (clean and watermarked) and record the first-stage distance for each. As described in \citet{arabi2024hidden}, this distance serves as the basis for watermark detection. We plot the histogram of the first-stage distance for clean and watermarked images, respectively, as shown in \cref{Fig App: First-stage distance}. 
We observe that thresholding the first-stage distance produced by the $\text{WIND}_{\text{inpainting}}$ decoder is insufficient for reliably distinguishing between clean and watermarked images. Consequently, we do not include $\text{WIND}_{\text{inpainting}}$ in our comparison.  

\end{document}